\documentclass[10pt,aps,prd,twocolumn,preprintnumbers,showpacs,nofootinbib,notitlepage,longbibliography,superscriptaddress]{revtex4-1}
\usepackage{amsmath, amssymb,braket}
\usepackage{wasysym} 
\usepackage{color}
\usepackage{float}
\usepackage{graphicx}
\usepackage{subfigure}
\usepackage{natbib}
\usepackage{tikz}
\usepackage[compat=1.1.0]{tikz-feynman}
\usepackage[colorlinks=true,citecolor=blue,urlcolor=blue]{hyperref}
\usepackage{xspace}
\usepackage{adjustbox}

\newcommand\Msun{\text{M}_{\astrosun}} 

\newcommand{\n}{\nonumber \\}
\newcommand{\dbar}{d\hspace*{-0.08em}\bar{}\hspace*{0.1em}}
\newcommand{\thesan}{\textsc{thesan}\xspace}

\newcommand\orcid[1]{\href{http://orcid.org/#1}{\adjustbox{trim={-.15\width} {0\height} {-.15\width} {0\height},clip}{\includegraphics[height=10pt]{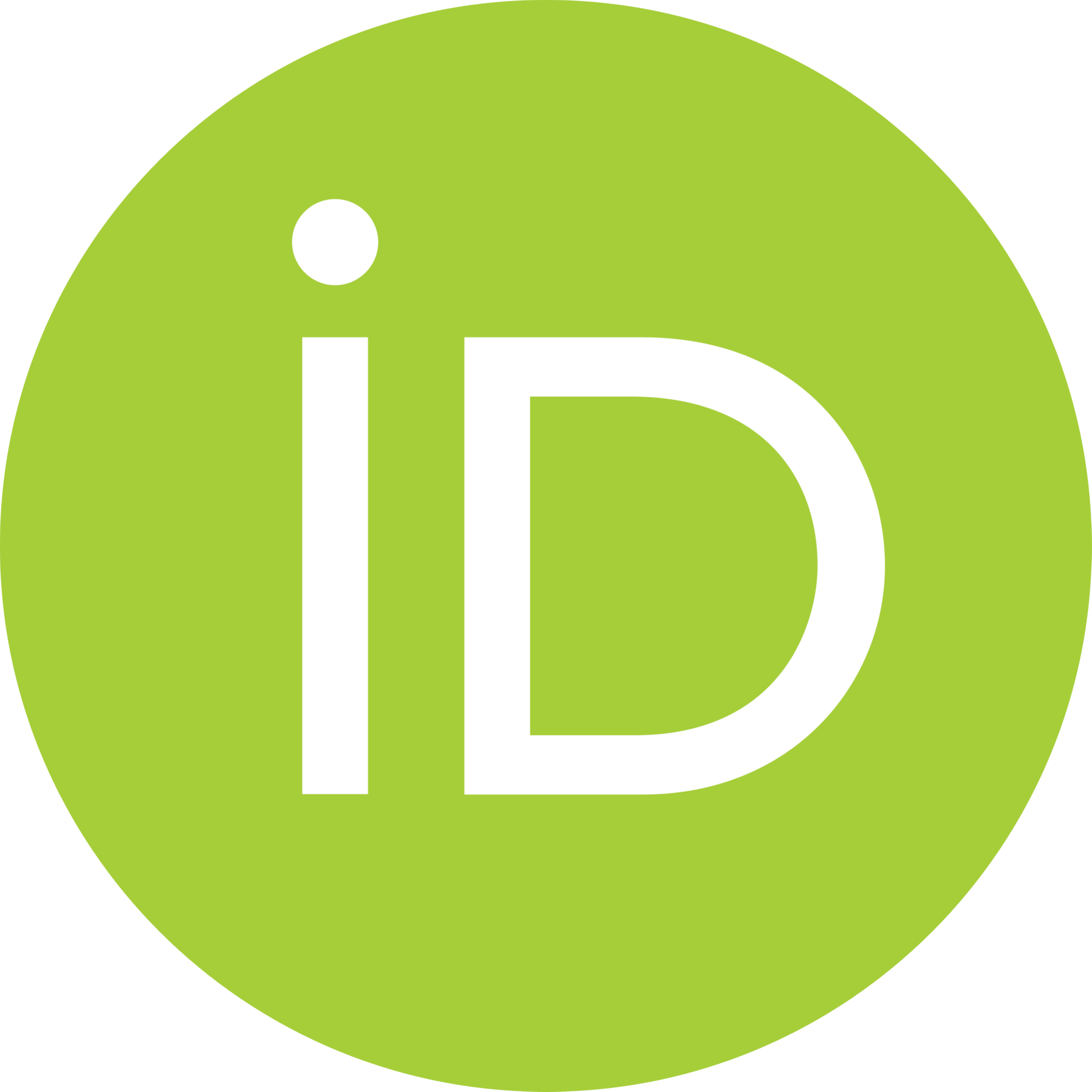}}}}

\begin{document}

\preprint{MIT-CTP/5418}

\title{An Effective Bias Expansion for 21 cm Cosmology in Redshift Space}

\author{Wenzer Qin\orcid{0000-0001-7849-6585}}
\affiliation{Center for Theoretical Physics, Massachusetts Institute of Technology, Cambridge, Massachusetts 02139, USA}

\author{Katelin Schutz\orcid{0000-0003-4812-5358}}
\thanks{NASA Hubble Fellowship Program Einstein Postdoctoral Fellow}
\affiliation{Center for Theoretical Physics, Massachusetts Institute of Technology, Cambridge, Massachusetts 02139, USA}
\affiliation{Department of Physics \& McGill Space Institute, McGill University, Montr\'{e}al, QC H3A 2T8, Canada}

\author{Aaron Smith\orcid{0000-0002-2838-9033}}
\thanks{NASA Hubble Fellowship Program Einstein Postdoctoral Fellow}
\affiliation{Center for Astrophysics $\vert$ Harvard \& Smithsonian, 60 Garden Street, Cambridge, MA 02138, USA}
\affiliation{MIT Kavli Institute for Astrophysics and Space Research, Massachusetts Institute of Technology, Cambridge, MA 02139, USA}

\author{\\Enrico Garaldi\orcid{0000-0002-6021-7020}}
\affiliation{Max-Planck Institute for Astrophysics, Karl-Schwarzschild-Str. 1, D-85741 Garching, Germany}

\author{Rahul Kannan\orcid{0000-0001-6092-2187}}
\affiliation{Center for Astrophysics $\vert$ Harvard \& Smithsonian, 60 Garden Street, Cambridge, MA 02138, USA}

\author{Tracy R. Slatyer\orcid{0000-0001-9699-9047}}
\affiliation{Center for Theoretical Physics, Massachusetts Institute of Technology, Cambridge, Massachusetts 02139, USA}

\author{Mark Vogelsberger\orcid{0000-0001-8593-7692}}
\affiliation{MIT Kavli Institute for Astrophysics and Space Research, Massachusetts Institute of Technology, Cambridge, MA 02139, USA}

\begin{abstract}
	\noindent 
	A near-future detection of the 21\,cm signal from the epoch of reionization will provide unique opportunities to probe the underlying cosmology, provided that such cosmological information can be extracted with precision. 
	To this end, we further develop  effective field theory (EFT) inspired techniques for the 21\,cm brightness temperature field during the epoch of reionization, incorporating renormalized bias and a treatment of redshift space distortions. 
	Notably, we confirm that in redshift space, measures of the 21\,cm brightness, e.g the power spectrum, should have irreducible contributions that lack a bias coefficient and therefore contain direct, astrophysics-free information about the cosmological density field; in this work, we study this effect beyond linear order.
	To validate our theoretical treatment, we fit the predicted EFT Fourier-space shapes to the \thesan suite of hydrodynamical simulations of reionization \emph{at the field level}, where the considerable number of modes prevents overfitting. 
	We find agreement at the level of a few percent between the 21\,cm power spectrum from the EFT fits and simulations over the wavenumber range $k \lesssim 0.8$ h/Mpc and neutral fraction $x_\mathrm{HI} \gtrsim 0.4$, which is imminently measurable by the Hydrogen Epoch of Reionization Array (HERA) and future experiments. 
	The ability of the EFT to describe the 21\,cm signal extends to simulations that have different astrophysical prescriptions for reionization as well as simulations with interacting dark matter. 
\end{abstract}

\maketitle

\section{Introduction}
\label{sec:intro}

The 21\,cm transition of neutral hydrogen provides a promising avenue for mapping out large scale structure (LSS) and testing cosmological theories at redshifts where there are few or no other detectable luminous tracers of the underlying matter field. 
Most empirical cosmological information either comes from measurements of the cosmic microwave background (CMB), which was emitted around the time of recombination $z \approx 1100$, or from surveys of tracers like galaxies at lower redshifts. 
To better understand how structure in our universe evolved at intermediate redshifts, we need observations of the diffuse neutral hydrogen gas from immediately after recombination through to the Epoch of Reionization (EoR).

Several experiments are already actively attempting to map the cosmological 21\,cm signal from the EoR, both at the level of the global signal (monopole) using experiments like EDGES~\cite{Monsalve:2016xbk}, LEDA~\cite{2018MNRAS.478.4193P}, PRI$^Z$M~\cite{2019JAI.....850004P}, and SARAS~\cite{Singh:2017syr}, as well as the fluctuations in the 21\,cm signal using interferometric experiements like PAPER~\cite{2010AJ....139.1468P}, the MWA ~\cite{2013PASA...30....7T,2013PASA...30...31B}, LOFAR~\cite{2013A&A...556A...2V}, HERA~\cite{DeBoer:2016tnn}, and the upcoming Square Kilometre Array (SKA)~\cite{Bull:2018lat}; there are also a number of post-reionization intensity mapping efforts such as CHIME~\cite{2014SPIE.9145E..22B}, HIRAX~\cite{Newburgh:2016mwi}, and CHORD~\cite{2019clrp.2020...28V}.
There has already been a tentative detection of the global signal from cosmic dawn in the form of a deep absorption trough at $z\sim 17$ made by the EDGES collaboration~\cite{Bowman:2018yin}, although this interpretation is in strong tension with observations from SARAS~3~\cite{Singh:2022}.
Further study of this feature is a major goal of 21\,cm experiments moving forward, while at the same time there is a push towards measuring the 21\,cm power spectrum from the EoR and eventually performing full tomographic mapping.

One complication of measuring the 21\,cm power spectrum is redshift space distortions (RSDs), which are contributions to the observed redshift that arise due to the peculiar velocities of neutral hydrogen rather than Hubble expansion. 
In other words, using the redshift of an observed line emission to infer a distance without accounting for line-of-sight peculiar velocities will yield the wrong distance. 
One can only directly measure distances in this illusory ``redshift space'' since there is no other independent way of inferring the peculiar velocity of the gas; 
thus, we are faced with the problem of extracting information about real space cosmology from redshift space observables. 
This is particularly relevant for interferometric measurements of the 21\,cm EoR signal, since substantial foregrounds that contaminate the signal lie in a ``wedge'' in $k_\parallel$ vs. $k_\perp$, where $k_\parallel$ and $k_\perp$ denote the line-of-sight and transverse components of Fourier modes, respectively. 
Modes with even moderate projections onto the $k_\perp$ direction will be within the foreground wedge whereas modes with larger projections onto the line-of-sight direction will be less contaminated by foregrounds~\cite{2010ApJ...724..526D,2012ApJ...745..176V,2012ApJ...752..137M,2012ApJ...756..165P,Trott:2012md,2013ApJ...768L..36P,2013ApJ...770..156H,Thyagarajan:2013eka,Liu:2014bba,Liu:2014yxa,2015ApJ...804...14T,2015ApJ...807L..28T,2016ApJ...833..242L,2016MNRAS.458.2928C,2018MNRAS.476.3051A}. 
In addition to evading foregrounds, from an instrumental perspective the high-$k$ modes that are nearly parallel to the line of sight are more readily observable due to the ease of attaining high spectral resolution as opposed to angular resolution. 
Experiments like HERA~\cite{DeBoer:2016tnn} therefore predominantly observe modes that are nearly parallel to the line of sight, and these are precisely the modes that will be most affected by RSDs~\cite{2016MNRAS.456...66J}.
For useful reviews on 21\,cm foreground mitigation, see Refs.~\cite{2019arXiv190912369C} and \cite{2020PASP..132f2001L}.

In this paper, we parametrize the effects of RSDs on the 21\,cm field using techniques inspired by effective field theory (EFT)~\cite{Baumann:2010tm,Carrasco:2012cv}. 
In recent years, EFT techniques have become a powerful tool for studying large scale structure~\cite{Carrasco:2013sva,Carrasco:2013mua,Carroll:2013oxa,Senatore:2014via,Baldauf:2015zga,Foreman:2015lca,Baldauf:2015aha,Cataneo:2016suz,Lewandowski:2017kes,Konstandin:2019bay,DAmico:2019fhj,Ivanov:2019pdj}. 
As structure formation progresses, nonlinear effects at a given scale become increasingly important; in other words, while density perturbations in the recombination epoch can be accurately described purely by linear theory, perturbations in the EoR cannot. With EFT techniques, one can systematically treat mildly nonlinear effects to increasingly high accuracy, up to some cutoff scale where structure formation becomes fully nonlinear. 
More specifically, we use EFT-inspired methods to treat the feedback of small-scale nonlinear effects on the larger scales of interest; this procedure is analogous to renormalization~\cite{Assassi:2014fva}.

\begin{figure*}
    \begin{tabular}{cc}
    \includegraphics[width=0.35\textwidth]{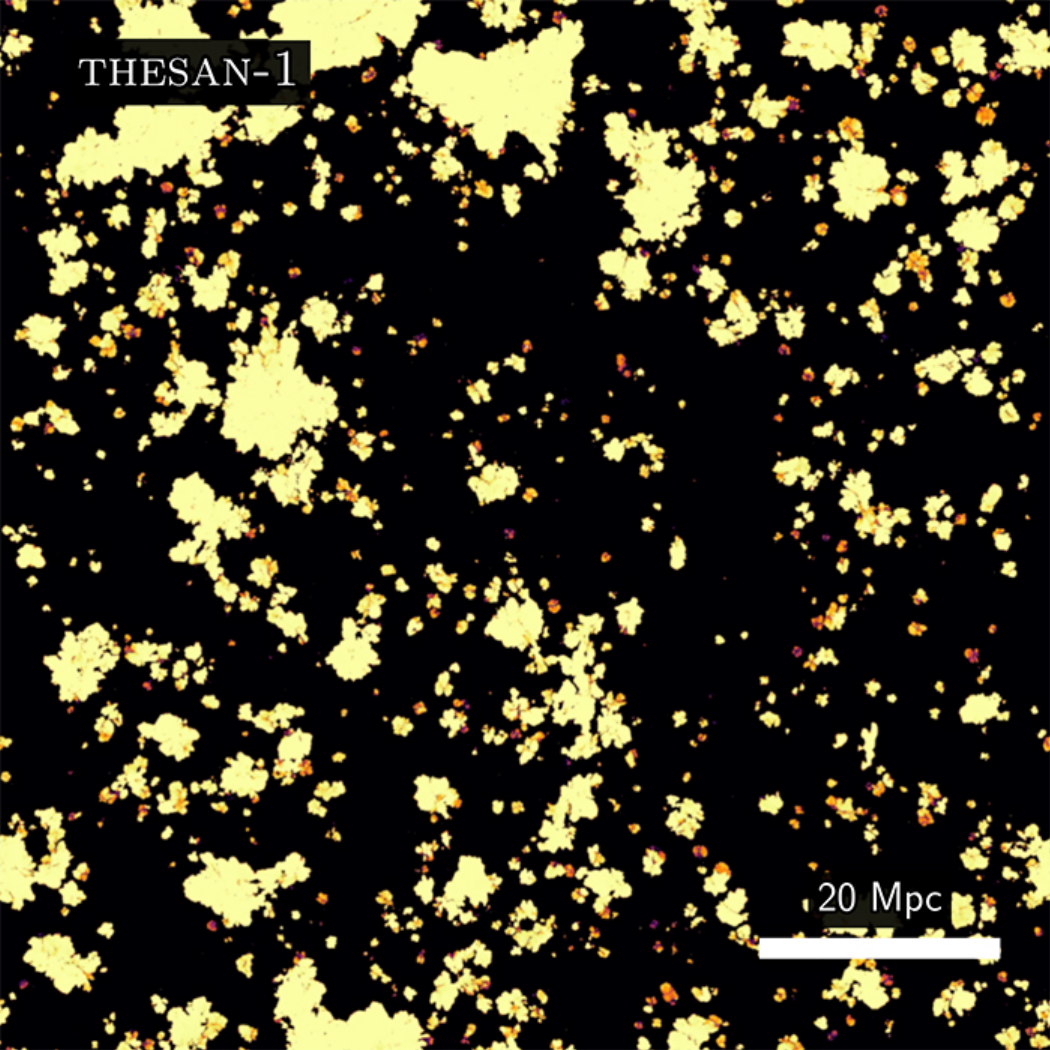} & \quad \includegraphics[width=0.35\textwidth]{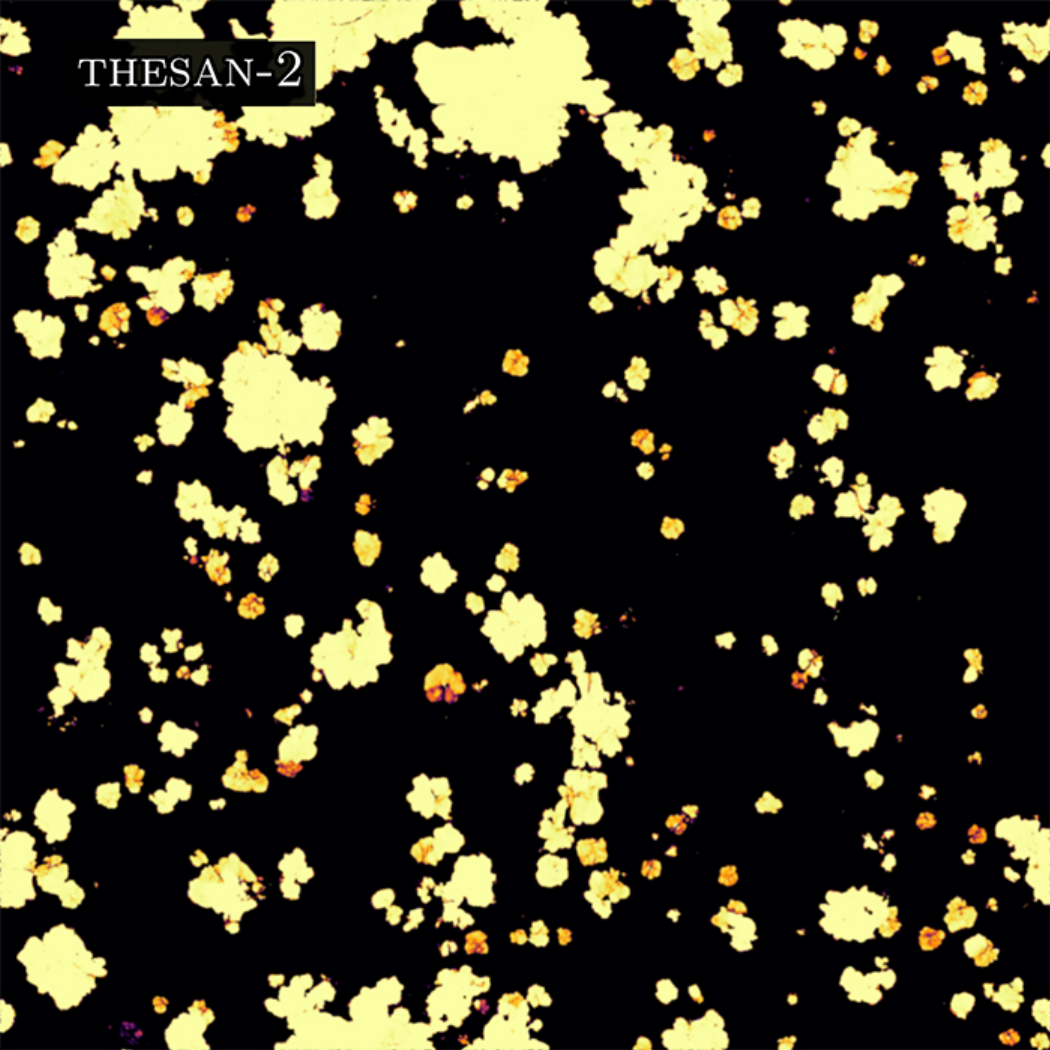} \\ \\
    \includegraphics[width=0.35\textwidth]{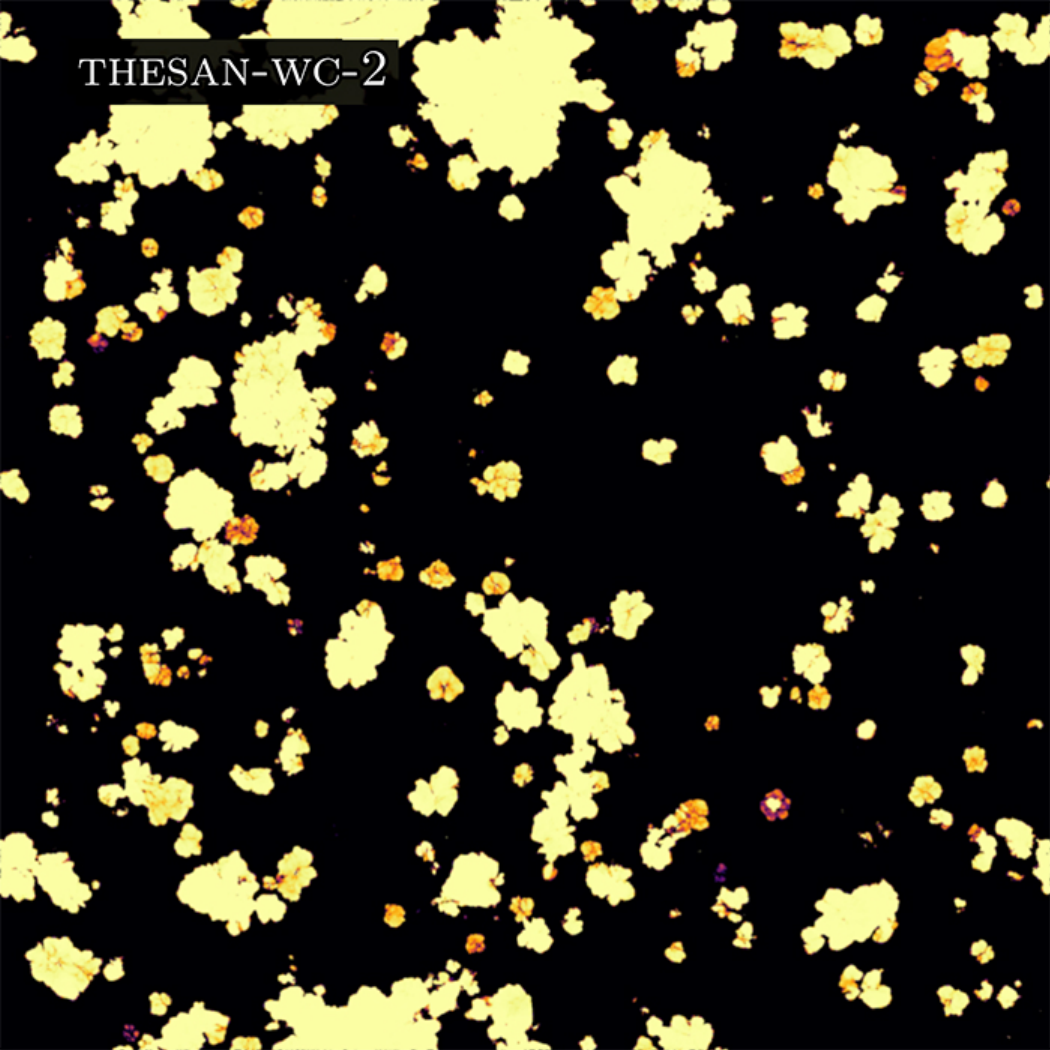} & \quad \includegraphics[width=0.35\textwidth]{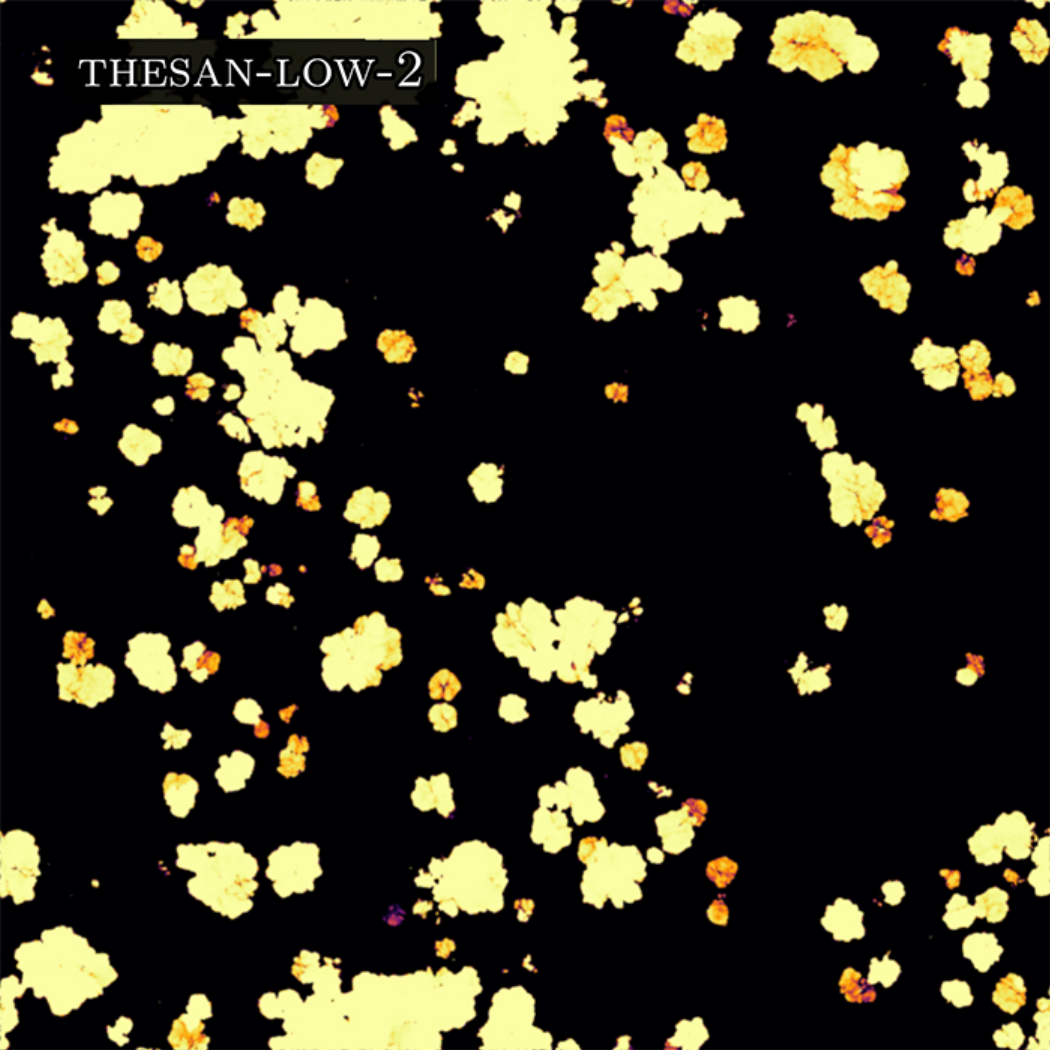} \\ \\
    \includegraphics[width=0.35\textwidth]{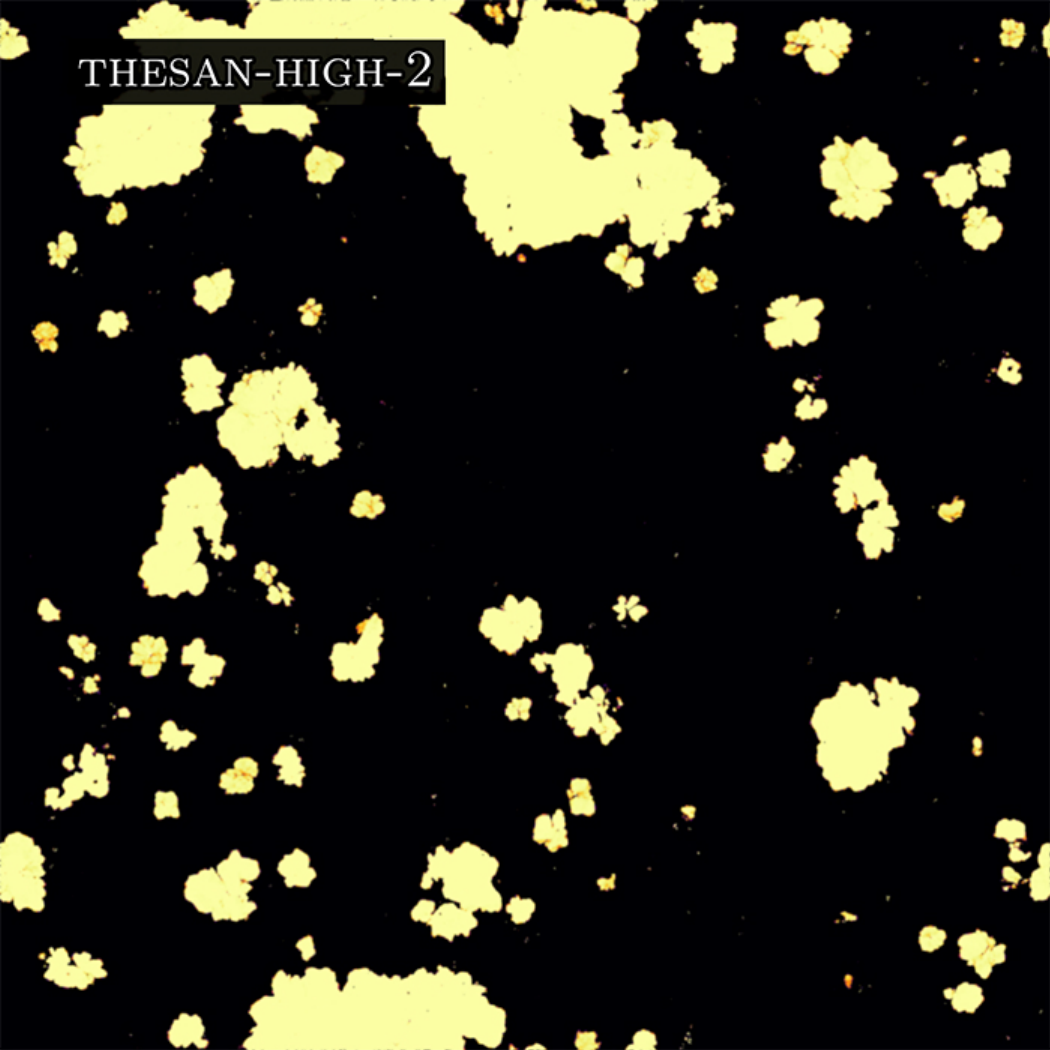} & \quad \includegraphics[width=0.35\textwidth]{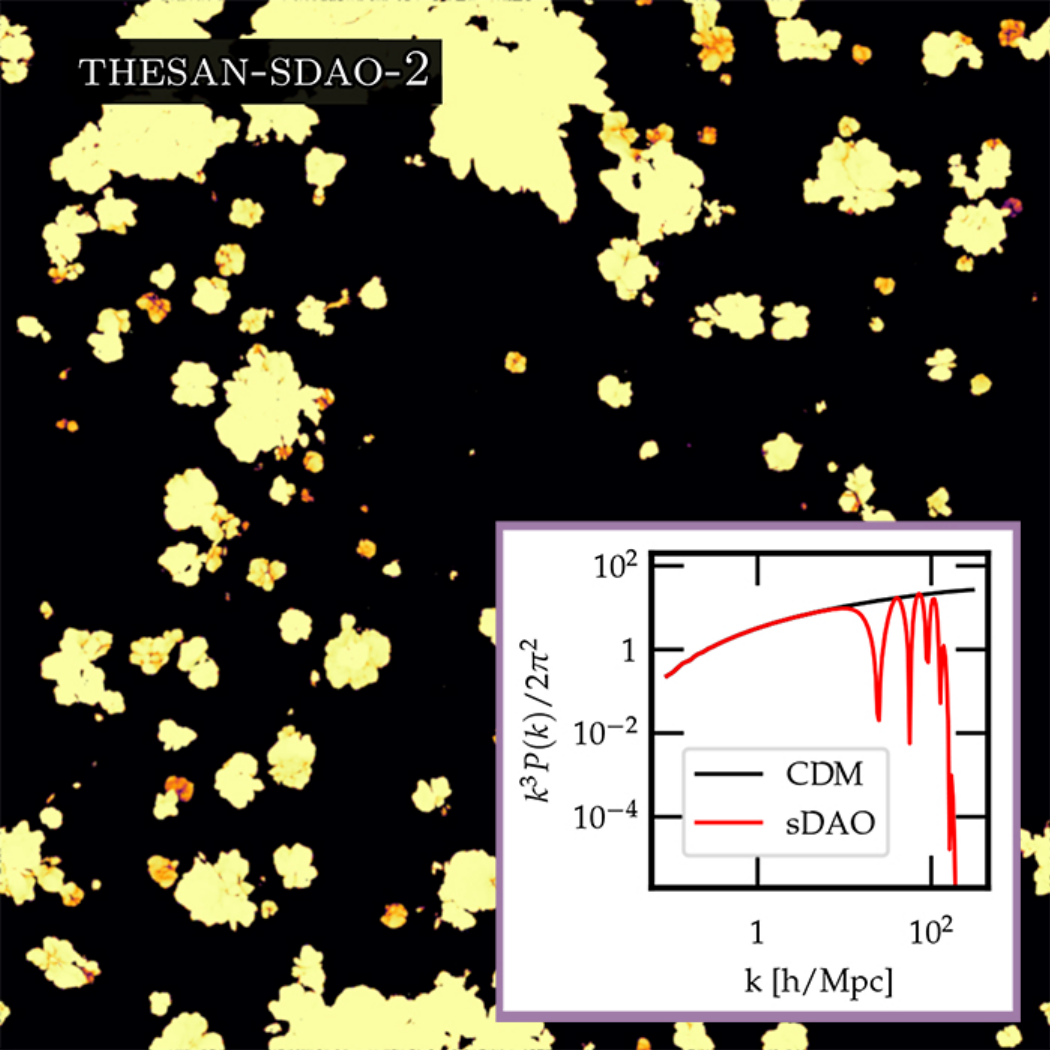}
    \end{tabular}
    \caption{Maps of the ionized bubble distribution in the different \thesan simulations, at $x_\mathrm{HI} \sim 0.7$.
    The bubbles are projected from a thick `slice' of the simulation spanning 8\% of the box volume.
    \thesan-1 is the highest resolution simulation in the \thesan suite.
    \thesan-2 is a medium resolution simulation that is otherwise the same as \thesan-1; \thesan-\textsc{wc}-2 has a slightly higher escape fraction to compensate for the lower star formation in \thesan-2 compared to \thesan-1.
    \thesan-\textsc{low}-2 is the same as \thesan-2 except that only halos below $10^{10}\,\Msun$ contribute to reionization, whereas in \thesan-\textsc{high}-2, only halos above $10^{10}\,\Msun$ contribute.
    \thesan-\textsc{sdao}-2 is the same as \thesan-2 but uses a non-standard dark matter model that effectively cuts off the linear matter power spectrum at small scales.
    The inset plot shows the linear matter power spectrum for cold dark matter and the dark acoustic oscillation model.
    As expected, \thesan-\textsc{high}-2 exhibits the largest ionized bubbles.
    Small ionized bubbles are also less abundant in the \thesan-\textsc{high}-2 and \thesan-\textsc{sdao}-2 simulations, compared to the others.}
    \label{fig:slices}
\end{figure*}

The application of EFT-inspired techniques to the EoR 21\,cm intensity field has only been studied relatively recently, as it was previously thought that the 21\,cm signal was nonperturbative in the wavenumber range probed by telescopes due to the presence of large ionized structures. 
To date, most of the theoretical analysis of the 21\,cm signal has been driven by computationally expensive radiative transfer simulations~\cite{gnedin1997reionization,ciardi2003simulating,McQuinn:2006et, iliev2006simulating,trac2007radiative,gnedin2014cosmic,pawlik2017aurora}, or by semi-analytic models such as 21\textsc{cmfast}~\cite{2011MNRAS.411..955M,2020JOSS....5.2582M} that can survey a wide range of theories of reionization with $\mathcal{O}(10\%)$ level agreement with simulation~\cite{zahn2007simulations,zahn2011comparison,majumdar2014use} (although the level of agreement depends sensitively on the ability of simulations to resolve self-shielded Lyman limit systems~\cite{kaurov2016cosmic}).
There have also been studies using phenomenological models that model the distribution of bubble sizes~\cite{2018ApJ...860...55R,2019ApJ...876...56R,Mirocha:2022}, parameterized models tuned to radiative transfer simulations~\cite{2013ApJ...776...81B}, models that match to a given mass-weighted ionization fraction~\cite{2022ApJ...927..186T}, and hybrid numerical methods that simulate the distribution of the first stars~\cite{2012Natur.487...70V,2014MNRAS.437L..36F}.

However, perturbative methods have gradually been developed with increasing success to study the process of reionization.
Linear perturbation theory can qualitatively reproduce many of the features of the EoR~\cite{2007MNRAS.375..324Z,2015PhRvD..91h3015M}, and theories including quadratic bias can match semi-analytic models to the level of tens of percent on very large scales~\cite{2019MNRAS.487.3050H}.
Ref.~\cite{McQuinn:2018zwa} pioneered the use of an effective bias expansion together with large-scale reionization simulations to show that the signal is in fact only mildly nonlinear on observable scales and that the field can be described accurately in real space with a small number of free parameters. 

In this paper, we extend the perturbative description of the 21\,cm signal to include RSDs, the effects of which have previously been encapsulated in EFT treatments of the density field and halos~\cite{Senatore:2014vja,Lewandowski:2015ziq,Perko:2016puo}. 
In particular, we find that the RSDs give rise to terms that are not multiplied by any bias coefficients; therefore, the contribution of these terms to observable quantities is fixed and does not add any degrees of freedom when fitting to measurements or simulations.
Previous works have studied these terms using linear theory, and found that these terms can enhance the power spectrum up to a factor of $\sim 2$~\cite{Barkana:2004zy} and that the size of the resulting anisotropies varies with redshift~\cite{2015PhRvL.114j1303F}.
Moreover, these contributions contain information about the underlying cosmological density field that is free from astrophysical influence. We test the validity of our theoretical approach using \thesan, a suite of state-of-the-art radiation hydrodynamic simulations~\cite{thesan1, thesan2, thesan3, thesan4, thesan5, thesan6}.
Figure \ref{fig:slices} shows example slices from the simulations included in the suite.

All distances in this paper are reported in terms of comoving units.
Readers mainly interested in the results of fitting the effective field theory expansion to simulations may choose to skip Section~\ref{sec:review}, which reviews perturbative methods for studying cosmological density fields, including standard perturbation theory (SPT), EFT, and the renormalization of local composite operators that appear in bias expansions.
In Section~\ref{sec:21cm}, we apply these perturbative techniques to the 21\,cm brightness temperature in redshift space. In Section~\ref{sec:thesan}, we introduce the \thesan simulations and describe our method for fitting the coefficients in the theory expansion to the simulations, which is done at the level of the cosmological \textit{fields}, instead of directly to the power spectrum to mitigate the possibility of overfitting. 
Section~\ref{sec:discussion} elaborates on the physical interpretation of the bias parameters and compares the fit parameters for simulations run with different physics, including a simulation with interacting dark matter that exhibits strong dark acoustic oscillations (sDAOs). 
Finally, we summarize our findings and outline some future directions in Section~\ref{sec:conclusion}.

\section{Review of Cosmological perturbation theories}
\label{sec:review}

In this section, we review results from SPT and the EFT of LSS, as well as effective renormalization in bias expansions. 
We begin by deriving the equations of motion for density and velocity perturbations and showing the perturbative solutions in SPT. 
We then review some results from EFT and introduce a diagrammatic language to help organize calculations involving higher-order terms and composite operators. 
See also Ref.~\cite{Desjacques:2016bnm} for a comprehensive review of cosmological perturbation theory, particularly in the context of galaxy bias.

\subsection{Standard Perturbation Theory}
\label{sec:spt}

Given a phase space distribution of collisionless particles $f (\tau, \boldsymbol{x}, \boldsymbol{p})$, the Boltzmann equation in an expanding universe is~\cite{Dodelson:2003ft}
\begin{equation}
    \frac{\mathrm{d}f}{\mathrm{d}t} = \frac{1}{a} \frac{\partial f}{\partial \tau} + \frac{\boldsymbol{p}}{a^2 m} \cdot \frac{\partial f}{\partial \boldsymbol{x}} - m \frac{\partial f}{\partial \boldsymbol{p}} \cdot \frac{\partial \phi}{\partial \boldsymbol{x}} = 0 .
\end{equation}
Here, $t$ denotes cosmic time, and is related to conformal time $\tau$ via $dt = a d\tau$, where $a$ is the scale factor, 
$\boldsymbol{x}$ and $\boldsymbol{p}$ are the comoving positions and momenta, and $\phi$ is the gravitational potential.
In this and following equations, we use boldface type to represent vectors; however, we occasionally use Einstein index notation to avoid ambiguities.

The first three moments of $f (\tau, \boldsymbol{x}, \boldsymbol{p})$ correspond to the comoving mass density, momentum density, and velocity dispersion:
\begin{align}
    \rho (\tau, \boldsymbol{x}) &\equiv m \int \dbar^3 p \, f (\tau, \boldsymbol{x}, \boldsymbol{p}) , \\
    \boldsymbol{\pi} (\tau, \boldsymbol{x}) &\equiv \int \dbar^3 p \, f (\tau, \boldsymbol{x}, \boldsymbol{p}) \, \boldsymbol{p} , \\
    \sigma^{ij} (\tau, \boldsymbol{x}) &\equiv \frac{1}{m^2} \int \dbar^3 p \, f (\tau, \boldsymbol{x}, \boldsymbol{p}) \, p^i p^j - \frac{\pi^i \pi^j}{m\rho} .
\end{align}
In these definitions, we denote $\dbar^3 p = d^3 p / (2\pi)^3$.
The first two moments of the Boltzmann equation correspond to the continuity equation
\begin{equation}
    0 = \partial_\tau \rho + \frac{1}{a} \boldsymbol{\nabla} \cdot \boldsymbol{\pi}
    \label{eqn:continuity}
\end{equation}
and the Euler equation
\begin{equation}
    0 = \partial_\tau \pi_i + \frac{1}{a} \partial_j \left( \frac{\pi_i \pi_j}{\rho} \right) + a \rho \nabla_i \phi .
    \label{eqn:euler}
\end{equation}
In the present formulation, the fluid equations contain no terms corresponding to shear forces, viscosity, or heat conduction; our collisionless particles therefore constitute a perfect fluid.

We solve these fluid equations perturbatively, defining $\delta = \rho/\bar{\rho} - 1$, where $\bar{\rho}$ is the mean density, and noting that the momentum density can be rewritten in terms of the physical peculiar velocity $\boldsymbol{v}$ as $\boldsymbol{\pi} = \rho a \boldsymbol{v}$; this is not the velocity of individual particles, but the bulk velocity of the field, i.e. averaged velocity of the particles in a region. 
Since the mean velocity of a homogeneous universe vanishes, $\boldsymbol{v}$ is perturbatively small. 
In terms of $\delta$ and $\boldsymbol{v}$, the continuity and Euler equations become
\begin{align}
    0 &= \partial_\tau \delta + \boldsymbol{\nabla} \cdot [ (1+\delta) \boldsymbol{v}] \\
    0 &= \partial_\tau \boldsymbol{v} + \mathcal{H} \boldsymbol{v} + (\boldsymbol{v} \cdot \boldsymbol{\nabla}) \boldsymbol{v} + \boldsymbol{\nabla} \phi
\end{align}
Here, $\mathcal{H} = \partial_\tau a / a$ is the conformal Hubble parameter.
We also include the Poisson equation in comoving coordinates as an equation of motion; since we are dealing with scales much smaller than the Hubble length, gravity can be treated as Newtonian:
\begin{equation}
    \partial^2 \phi = \frac{3}{2} \mathcal{H}^2 \Omega_m \delta.
\end{equation}
\begin{widetext}
Above, $\Omega_m$ is the mass density in units of the critical density.
We hereafter set $\Omega_m=1$ since reionization occurs deep in the matter-dominated era.
The velocity can be further decomposed in terms of its divergence, $\theta = \boldsymbol{\nabla} \cdot \boldsymbol{v}$, and curl or vorticity, $\boldsymbol{\omega} = \boldsymbol{\nabla} \times \boldsymbol{v}$.
However, at leading order in SPT, any initial vorticity decays linearly with the expansion of the Universe; therefore, we neglect the contribution to the velocity field coming from $\boldsymbol{\omega}$.
\footnote{At higher order in SPT, there can be growing vorticity modes; however, the sources always contain powers of the vorticity at linear order, and are therefore still suppressed relative to the growing modes of $\delta$ and $\theta$.
Vorticity can also matter in EFT at third order, because the stress tensor and heat conduction terms source a non-decaying contribution~\cite{Bertolini:2016bmt}.}
With this velocity decomposition, in Fourier space the continuity and Euler equations are
\begin{align}
	\partial_\tau \delta_{\boldsymbol{k}} + \theta_{\boldsymbol{k}} &= - \int \dbar^3 q \, \frac{\boldsymbol{q} \cdot \boldsymbol{k}}{\boldsymbol{q}^2} \, \theta_{\boldsymbol{q}} \delta_{(\boldsymbol{k}-\boldsymbol{q})} , \label{eqn:S_a} \\
	\partial_\tau \theta_{\boldsymbol{k}} + \mathcal{H} \theta_{\boldsymbol{k}} + \frac{3}{2} \mathcal{H}^2 \delta_{\boldsymbol{k}} &= - \int \dbar^3 q \, \left( \frac{k^2 [\boldsymbol{q} \cdot (\boldsymbol{k} - \boldsymbol{q})]}{2 q^2 (\boldsymbol{k} - \boldsymbol{q})^2} \right) \theta_{\boldsymbol{q}} \theta_{(\boldsymbol{k}-\boldsymbol{q})} . \label{eqn:S_b}
\end{align}
Above, we've used bold subscripts to denote Fourier transformed quantities, e.g. $\delta_{\boldsymbol{k}} = \int d^3 x \, \delta (\boldsymbol{x}) e^{-i \boldsymbol{k} \cdot \boldsymbol{x}}$.
It will also be useful to convert the time derivatives into derivatives with respect to scale factor using $\partial_\tau = \mathcal{H} a \partial_a$ and the fact that $\mathcal{H} \propto 1 / \sqrt{a}$ during matter domination.

The right hand sides of Eqns. \eqref{eqn:S_a} and \eqref{eqn:S_b} mix the $\delta$ and $\theta$ modes.
To solve these coupled fluid equations, one can adopt the perturbative ansatz
\begin{gather}
    \delta_{\boldsymbol{k}} (\tau) = \sum_{n=1}^\infty D^n (\tau) \delta^{(n)}_{\boldsymbol{k}} , \\
    \theta_{\boldsymbol{k}} (\tau) = - \mathcal{H}(\tau) f(\tau) \sum_{n=1}^\infty D^n (\tau) \theta^{(n)}_{\boldsymbol{k}}
\end{gather}
where $\delta^{(n)}$ and $\theta^{(n)}$ are $\mathcal{O}(\delta^{(1)})^n$ and where $D(\tau)$ and $f(\tau) = d \ln D(\tau) / \mathcal{H} d\tau $ are the linear and logarithmic growth functions. Since reionization occurs deep in the matter-dominated epoch, we set $D(\tau) = a (\tau)$ and $f(\tau) = 1$ in this work; to include the effect of a dark energy component, one can substitute the appropriate growth factors~\cite{Bernardeau:2001qr}. Given the form of the ansatz, the solution to Eqns. \eqref{eqn:S_a} and \eqref{eqn:S_b} can be expressed as
\begin{align}
	\delta^{(n)}_{\boldsymbol{k}} &= \int \dbar^3 q_1 \dots \int \dbar^3 q_n \, (2\pi)^3 \delta^D \left( \boldsymbol{k} - \sum_{i=1}^n \boldsymbol{q}_i \right) F_n (\boldsymbol{q}_1, \dots, \boldsymbol{q}_n) \delta^{(1)}_{\boldsymbol{q}_1} \dots \delta^{(1)}_{\boldsymbol{q}_n} , \label{eqn:dn} \\
	\theta^{(n)}_{\boldsymbol{k}} &= \int \dbar^3 q_1 \dots \int \dbar^3 q_n \, (2\pi)^3 \delta^D \left( \boldsymbol{k} - \sum_{i=1}^n \boldsymbol{q}_i \right) G_n (\boldsymbol{q}_1, \dots, \boldsymbol{q}_n) \delta^{(1)}_{\boldsymbol{q}_1} \dots \delta^{(n)}_{\boldsymbol{q}_n}  \label{eqn:tn}
\end{align}
where the mode coupling kernels $F_n$ and $G_n$ have well-known recursion relations~\cite{Goroff:1986ep,Jain:1993jh,Bernardeau:2001qr}.
The first few kernels, symmetrized over permutations of the momenta, are
\begin{gather}
    F_1 = G_1 = 1 , \\
    F_2(\boldsymbol{q}_1, \boldsymbol{q}_2) = \frac{5}{7} + \frac{2}{7} \frac{(\boldsymbol{q}_1 \cdot \boldsymbol{q}_2)^2}{\boldsymbol{q}_1^2 \boldsymbol{q}_2^2} + \frac{\boldsymbol{q}_1 \cdot \boldsymbol{q}_2}{2} \left( \frac{1}{\boldsymbol{q}_1^2} + \frac{1}{\boldsymbol{q}_2^2} \right) , \\
    G_2(\boldsymbol{\boldsymbol{q}}_1, \boldsymbol{\boldsymbol{q}}_2) = \frac{3}{7} + \frac{4}{7} \frac{(\boldsymbol{q}_1 \cdot \boldsymbol{q}_2)^2}{\boldsymbol{q}_1^2 \boldsymbol{q}_2^2} + \frac{\boldsymbol{q}_1 \cdot \boldsymbol{q}_2}{2} \left( \frac{1}{\boldsymbol{q}_1^2} + \frac{1}{\boldsymbol{q}_2^2} \right) .
\end{gather}
One can calculate correlation functions of these fields using a diagrammatic representation.
The diagram rules are:
\begin{enumerate}
    \item Each $\delta^{(n)}_{\boldsymbol{k}}$ and $\theta^{(n)}_{\boldsymbol{k}}$ corresponds to a vertex with one external leg of wavenumber $\boldsymbol{k}$ and $n$ internal legs representing the factors of $\delta^{(1)}_{\boldsymbol{q}_i}$.
    The vertex couples the $n$ modes of the internal legs, and therefore corresponds to $F_n (\boldsymbol{q}_1, \dots, \boldsymbol{q}_n)$ or $G_n (\boldsymbol{q}_1, \dots, \boldsymbol{q}_n)$ depending on which field is involved. In analogy to conservation of momentum, wavenumber is conserved so each vertex also carries a factor of $(2\pi)^3 \delta^D \left( \boldsymbol{k} - \sum_{i=1}^n \boldsymbol{q}_i \right)$.
    We used filled dots to represent the density field and open dots to represent the velocity field.
    \begin{equation}
        \delta^{(n)}_{\boldsymbol{k}} \quad \rightarrow \quad
    	\begin{tikzpicture}[baseline=(current bounding box.center)]
    	\begin{feynman}
        	\vertex (i);
        	\vertex [right=1.5cm of i, scale=1.5, dot] (j) {};
        	
        	\vertex [above right=2cm of j] (a) {};
        	\vertex [below=0.5cm of a] (b) {};
        	\vertex [below right=2cm of j] (c) {};
        	
        	\vertex [below=1cm of b, scale=0.3, dot] (l) {};
        	\vertex [below=0.2cm of l, scale=0.3, dot] (m) {};
        	\vertex [below=0.2cm of m, scale=0.3, dot] (n) {};
        	
        	\diagram*{
        		(i) -- [edge label=\(\boldsymbol{k}\)] (j) ,
        		(j) -- [scalar, edge label=\(\boldsymbol{q_1}\)] {(a)},
        		(j) -- [scalar] {(b)},
        		(j) -- [scalar, edge label'=\(\boldsymbol{q_n}\)] {(c)}
        	};
    	\end{feynman}
    	\end{tikzpicture}
    	\quad = \quad (2\pi)^3 \delta^D \left( \boldsymbol{k} - \sum_{i=1}^n \boldsymbol{q}_i \right) F_n (\boldsymbol{q}_1, \dots, \boldsymbol{q}_n)
    \end{equation}
    \begin{equation}
        \theta^{(n)}_{\boldsymbol{k}} \quad \rightarrow \quad
    	\begin{tikzpicture}[baseline=(current bounding box.center)]
    	\begin{feynman}
        	\vertex (i);
        	\vertex [right=1.5cm of i, scale=1.5, empty dot] (j) {};
        	
        	\vertex [above right=2cm of j] (a) {};
        	\vertex [below=0.5cm of a] (b) {};
        	\vertex [below right=2cm of j] (c) {};
        	
        	\vertex [below=1cm of b, scale=0.3, dot] (l) {};
        	\vertex [below=0.2cm of l, scale=0.3, dot] (m) {};
        	\vertex [below=0.2cm of m, scale=0.3, dot] (n) {};
        	
        	\diagram*{
        		(i) -- [edge label=\(\boldsymbol{k}\)] (j) ,
        		(j) -- [scalar, edge label=\(\boldsymbol{q_1}\)] {(a)},
        		(j) -- [scalar] {(b)},
        		(j) -- [scalar, edge label'=\(\boldsymbol{q_n}\)] {(c)}
        	};
    	\end{feynman}
    	\end{tikzpicture}
    	\quad = \quad (2\pi)^3 \delta^D \left( \boldsymbol{k} - \sum_{i=1}^n \boldsymbol{q}_i \right) G_n (\boldsymbol{q}_1, \dots, \boldsymbol{q}_n)
    \end{equation}
    \item To compute a correlation function, draw all connected diagrams that can be made by contracting the internal $\delta^{(1)}$ legs. 
    Since we are using symmetrized $F_n$ and $G_n$ kernels, permuting the $\delta^{(1)}$ legs on each $\delta^{(n)}$ vertex will give rise to a symmetry factor of $n!$, and loops introduce additional combinatoric factors to prevent double-counting.
    \item For each internal leg carrying wavenumber $\boldsymbol{p}$, write down a factor of $P_L (\boldsymbol{p})$, the linear matter power spectrum. This is analogous to the propagator of the linear, ``free'' density fields of the internal legs, since the power spectrum is related to the two-point correlation function.
    \begin{equation}
        \begin{tikzpicture}
    	\begin{feynman}
    	\vertex [dot, scale=1.5] (i) {};
    	\vertex [right=2cm of i, empty dot, scale=1.5] (j) {};
    	\diagram*{
    		(i) --[scalar, momentum=\(\boldsymbol{p}\)] (j)
    	};
    	\end{feynman}
    	\end{tikzpicture}
    	\quad = \quad P_L (\boldsymbol{p})
    \end{equation}
    The vertices on the ends of the propagator can correspond to both $\delta$, both $\theta$, or one of each; the factor of $P_L (\boldsymbol{p})$ associated with the propagator is the same in any case, since $\delta^{(1)}$ and $\theta^{(1)}$ are spatially the same up to time-dependent factors.
    \item Integrate over the wavenumber $\boldsymbol{q}$ of each loop with $\int \dbar^3 q$.
\end{enumerate}
\end{widetext}

\subsection{Effective Field Theory}
\label{sec:eft}

For small-wavelength modes, perturbations will have collapsed enough to have become nonlinear and be outside of the regime of validity of the present perturbative theoretical treatment. 
These non-linearities can also affect large scales, since modes of different scales are coupled by the vertex kernels of Eqns.~\eqref{eqn:dn} and \eqref{eqn:tn} and since integrals over loops formally run over all wavenumbers.
This motivates introducing a smoothed version of the fields, where we convolve the densities or velocities with a windowing function $W_\Lambda$ of characteristic length scale $1/\Lambda$. 
We can apply this smoothing to the equations of motion e.g. Eqns. \eqref{eqn:continuity} and \eqref{eqn:euler}; however, smoothed composite operators can not be straightfowardly expressed as a product of smoothed fields, e.g. $(\delta v)_\text{smooth} \neq \delta_\text{smooth} \,v_\text{smooth}$. 
In order to express the equations of motion in terms of smoothed fields, one can express the smoothed composite operators as a product of smoothed fields after introducing additional correction terms~\cite{Baumann:2010tm,Carrasco:2012cv,Pajer:2013jj,Mercolli:2013bsa,Abolhasani:2015mra}. 
These terms are unknown \emph{a priori} but can be constructed from the bottom-up from all terms consistent with the symmetries (e.g. Galilean invariance). 
These terms take the form of an effective stress tensor for the long-wavelength fluid and the sensitivity to unknown behaviour of small-scale modes is parameterized as an effective speed of sound, viscosity, shear, etc. 
In other words, the smoothed density field is not a perfect fluid because of the feedback from small-scale modes.

The new terms in the effective stress tensor can be constructed order by order, by expanding the stress tensor in terms of convective time derivatives (co-moving with fluid elements) of local operators~\cite{Bertolini:2016bmt}.
As a result of this change to the equations of motion, we must modify the perturbative ansatz to include additional counterterms, which we denote by $\tilde{\delta}^{(n)}$ and $\tilde{\theta}^{(n)}$,
\begin{gather}
    \delta_{\boldsymbol{k}} (\tau) = \sum_{n=1}^\infty \left(  a(\tau)^n \delta^{(n)}_{\boldsymbol{k}} + \epsilon a(\tau)^{n+2} \tilde{\delta}^{(n)}_{\boldsymbol{k}} \right) \label{eqn:EFT_delta} \\
    \theta_{\boldsymbol{k}} (\tau) = - \mathcal{H}(\tau)  \sum_{n=1}^\infty \left( a(\tau)^n \theta^{(n)}_{\boldsymbol{k}} + \epsilon a(\tau)^{n+2} \tilde{\theta}^{(n)}_{\boldsymbol{k}} \right). \label{eqn:EFT_theta}
\end{gather}
Here, $\epsilon$ is a parameter that allows us to keep track of the EFT power counting. The counterterms come with an additional factor of $a (\tau)^2$ compared to the SPT terms because the EFT terms must have the same time dependence as loop contributions from SPT in order to correct them. 
The EFT kernels $\tilde{F}_n$ and $\tilde{G}_n$ are analogously defined relative to the SPT kernels as
\begin{widetext}
\begin{align}
	\tilde{\delta}^{(n)}_{\boldsymbol{k}} &= \int \dbar^3 q_1 \dots \int \dbar^3 q_n \, (2\pi)^3 \delta^D \left( \boldsymbol{k} - \sum_{i=1}^n q_i \right) \tilde{F}_n (\boldsymbol{q}_1, \dots, \boldsymbol{q}_n) \delta^{(1)}_{\boldsymbol{q}_1} \dots \delta^{(1)}_{\boldsymbol{q}_n} , \\
	\tilde{\theta}^{(n)}_{\boldsymbol{k}} &= \int \dbar^3 q_1 \dots \int \dbar^3 q_n \, (2\pi)^3 \delta^D \left( \boldsymbol{k} - \sum_{i=1}^n q_i \right) \tilde{G}_n (\boldsymbol{q}_1, \dots, \boldsymbol{q}_n) \delta^{(1)}_{\boldsymbol{q}_1} \dots \delta^{(1)}_{\boldsymbol{q}_n} .
\end{align}
\end{widetext}
The forms of the EFT kernels, $\tilde{F}_n$ and $\tilde{G}_n$, are derived in Ref.~\cite{Bertolini:2016bmt} and listed up to $n=3$. 
Correlation functions can then be computed and are robust to the effects of non-linearities affecting the results at the level of the fluid equations; i.e. the large perturbative scales are less affected by the uncertainties of small scale physics.

\subsection{Renormalized bias}
\label{sec:renorm}

Much of the formalism for cosmological effective field theories has been developed in the context of large scale structure and the matter density field.
However, EFT techniques can also be extended to study biased tracers of the matter field, such as galaxies and halos~\cite{2009JCAP...08..020M,Assassi:2014fva,Senatore:2014eva,Angulo:2015eqa,Fujita:2016dne,Perko:2016puo,Nadler:2017qto,Donath:2020abv}. 
Cosmological 21\,cm radiation is also a biased tracer of the underlying matter field on large scales.
When expressing the 21\,cm intensity field as a local bias expansion in terms of $\delta$, one needs to include all operators that respect homogeneity and isotropy; specifically, the theoretical 21\,cm field is built up only from operators that obey these symmetries, and should generically include contributions from all such operators,
\begin{align}
    (\delta_{21})_{\boldsymbol{k}} =& b_1 \delta_{\boldsymbol{k}} - b_{\nabla^2} k^2 \delta_{\boldsymbol{k}} \n
    &\quad + b_2 \left(\delta^2\right)_{\boldsymbol{k}} + b_{\mathcal{G}2} (\mathcal{G}_2)_{\boldsymbol{k}} + \cdots
    \label{eqn:bias}
\end{align}
Note that the momentum subscript denotes a Fourier transformation over the entire operator, e.g. $\left(\delta^2\right)_{\boldsymbol{k}} = \int \dbar^3 k \, \delta^2 (\boldsymbol{x}) e^{-i \boldsymbol{k} \cdot \boldsymbol{x}} \neq \left(\delta_{\boldsymbol{k}}\right)^2$.
In the above equation, $\mathcal{G}_2$ is the second Galileon or tidal operator, defined in configuration space as
\begin{equation}
    \mathcal{G}_2 = (\nabla_i \nabla_j \phi) (\nabla^i \nabla^j \phi) - (\nabla^2 \phi)^2 .
\end{equation}
The bias coefficients $b$ for the various operators in Eq.~\eqref{eqn:bias} are not known \emph{a priori} and must be determined from real or simulated data; in fact, the field on the left-hand side of Eqn.~\eqref{eqn:bias} can be replaced with any biased tracer of the underlying matter field, e.g. halos or galaxies, and the inferred coefficients will differ depending on the physics of the particular tracer in question.

From Eqn.~\eqref{eqn:bias}, we see that composite operators such as $\delta^2 (\boldsymbol{x})$ appear in the configuration space picture. 
Diagramatically, we represent this composite operator in Fourier space with the vertex
\begin{equation}
	\begin{tikzpicture}[baseline=(current bounding box.center)]
	\begin{feynman}
	\vertex (i);
	\vertex [right=1cm of i, small, blob] (a) {};
	\vertex [above right=1cm of a, dot] (b) {};
	\vertex [below right=1cm of a, dot] (c) {};
	\diagram*{
		(i) -- (a) ,
		(a) -- {(b),(c)}
	};
	\end{feynman}
	\end{tikzpicture}
\end{equation}
where the ``blob'' indicates a convolution in Fourier space and where the two legs ending in solid dots represent $\delta^{(n)}$ component fields entering the convolution.
Because these operators are local in configuration space, in Fourier space the convolution includes all wavenumbers.
Therefore, these composite operators contain contributions from small-scale modes that are non-linear, in analogy to the previous subsection. These non-linear contributions to the bias expansion are not removed by the EFT formalism described above, because the counterterms in Eqns. \eqref{eqn:EFT_delta} and \eqref{eqn:EFT_theta} only correct the non-linearities that affect the equations of motion for matter.
We follow the renormalization procedure in Ref.~\cite{Assassi:2014fva} to remove the small-scale or UV-dependence of composite operators order by order. 

To renormalize an operator $f$, we take correlation functions of $f$ with factors of the linear density field $\delta^{(1)}$ and add counterterms that cancel UV-sensitive loop contributions to these correlation functions in the zero-mode limit; this leaves only the tree-level (or zero-loop) contribution.
In other words, our renormalization condition is
\begin{align}
	\braket{[f_{\boldsymbol{k}}] \delta^{(1)}_{\boldsymbol{q}_1} \cdots \delta^{(1)}_{\boldsymbol{q}_n}} &= \braket{f_{\boldsymbol{k}} \delta^{(1)}_{\boldsymbol{q}_1} \cdots \delta^{(1)}_{\boldsymbol{q}_n}}_\mathrm{tree} \, \mathrm{for} \, \boldsymbol{q}_i = 0 , \, \forall \, i.
	\label{eqn:renorm_condition}
\end{align}
where the square brackets denote the renormalized operator $[f] = f + \sum_\mathcal{O} Z^f_\mathcal{O} \mathcal{O}$ such that the sum over all counterterm operators $\mathcal{O}$,  $\sum_\mathcal{O} Z^f_\mathcal{O} \mathcal{O}$ cancel the loop contributions. 
We evaluate the renormalization conditions at zero wavenumber, since this is the limit where the theory is most perturbative.
In evaluating loops, we only include diagrams where the loops connect multiple component fields of the convolution vertex; such diagrams are sometimes called ``one particle irreducible" or 1PI due to their similarity with such diagrams from quantum field theory~\cite{Assassi:2014fva,Abolhasani:2015mra}, but we stress that these definitions are not exactly the same.
The diagrams we include capture the additional mixing between small and large scale modes in the convolution that we are concerned with, as opposed to the mixing that arises from the equations of motion. We now make these definitions more explicit:
\begin{itemize}
    \item 1PI diagrams are diagrams that cannot be separated into two valid, disconnected diagrams by cutting a single internal line.
    The following graph is an example of a fully 1PI diagram.
    \begin{equation}
    \begin{tikzpicture}[baseline=(current bounding box.center)]
	\begin{feynman}
	\vertex (i);
	\vertex [right=1cm of i, small, blob] (a) {};
	\vertex [above right=1cm of a, dot] (b1) {};
	\vertex [right=1cm of a, dot] (b2) {};
	\vertex [below right=1cm of a, dot] (b3) {};
	\vertex [right=1cm of b1, dot] (c1) {};
	\vertex [right=1cm of c1] (d1);
	\vertex [right=1cm of b3, dot] (c2) {};
	\vertex [right=1cm of c2] (d2);
	\diagram*{
		(i) -- (a) ,
		(a) -- {(b1),(b2),(b3)},
		(b1) -- [scalar] (c1),
		(b1) -- [scalar] (b2),
		(c1) -- (d1),
		(b3) -- [scalar] (c2),
		(b3) -- [scalar] (b2),
		(c2) -- (d2),
	};
	\end{feynman}
	\end{tikzpicture}
	\nonumber
	\end{equation}
    Since the momenta in the external legs coming out of the composite operator are related via loops, this diagram involves mode mixing, so we include it in the renormalization procedure.
    Note that this example is a two-loop diagram, so we do not include this for calculating the one-loop power spectrum.
    
    \item We also include diagrams of the following type.
    \begin{equation}
    \begin{tikzpicture}[baseline=(current bounding box.center)]
	\begin{feynman}
	\vertex (i);
	\vertex [right=1cm of i, small, blob] (a) {};
	\vertex [above right=1cm of a, dot] (b1) {};
	\vertex [right=1cm of a, dot] (b2) {};
	\vertex [below right=1cm of a, dot] (b3) {};
	\vertex [right=1cm of b1, dot] (c1) {};
	\vertex [right=1cm of c1] (d1);
	\vertex [right=1cm of b3, dot] (c2) {};
	\vertex [right=1cm of c2] (d2);
	\diagram*{
		(i) -- (a) ,
		(a) -- {(b1),(b2),(b3)},
		(b1) -- [scalar] (c1),
		(b1) -- [scalar] (b2),
		(c1) -- (d1),
		(b3) -- [scalar] (c2),
		(c2) -- (d2),
	};
	\end{feynman}
	\end{tikzpicture}
	\nonumber
	\end{equation}
	This is not 1PI in the conventional sense, since it can be separated into two valid diagrams by cutting the bottom leg coming out of the convolution vertex.
	However, we still include it for the purpose of renormalization, since there is a loop that relates the momenta of the other two legs.
	Such diagrams have been termed ``partially 1PI" in the literature~\cite{Assassi:2014fva}.
    
    \item Below is an example of a diagram that is neither fully 1PI nor partially 1PI.
    \begin{equation}
    \begin{tikzpicture}[baseline=(current bounding box.center)]
	\begin{feynman}
	\vertex (i);
	\vertex [right=1cm of i, small, blob] (a) {};
	\vertex [above right=1cm of a, dot] (b1) {};
	\vertex [right=1cm of a, dot] (b2) {};
	\vertex [below right=1cm of a, dot] (b3) {};
	\vertex [right=1cm of b1, dot] (c1) {};
	\vertex [right=1cm of c1] (d1);
	\vertex [right=1cm of b2, dot] (c2) {};
	\vertex [right=1cm of c2] (d2);
	\vertex [right=1cm of b3, dot] (c3) {};
	\vertex [right=1cm of c3] (d3);
	\diagram*{
		(i) -- (a) ,
		(a) -- {(b1),(b2),(b3)},
		(b1) -- [scalar] (c1),
		(c1) -- (d1),
		(b2) -- [scalar] (c2),
		(c2) -- (d2),
		(b3) -- [scalar] (c3),
		(c3) -- (d3),
	};
	\draw[dashed] (b1) arc [start angle=270, end angle=-180, radius=0.3cm];
	\end{feynman}
	\end{tikzpicture}
	\nonumber
	\end{equation}
	We do not include this for bias renormalization, since the momenta running through the external legs of the composite operator vertex do not mix.
\end{itemize}

As an example, we show the calculation of the first few of counterterms for $\delta^2$. 
The first counterterm cancels UV sensitivity from the expectation value of $\delta^2$,
\begin{align}
    \braket{\left(\delta^2\right)_{\boldsymbol{k}}} \quad &= \quad
	\begin{tikzpicture}[baseline=(current bounding box.center)]
	\begin{feynman}
	\vertex (i);
	\vertex [right=1cm of i, small, blob] (a) {};
	\vertex [above right=1cm of a, dot] (b) {};
	\vertex [below right=1cm of a, dot] (c) {};
	\diagram*{
		(i) -- [momentum=\(\boldsymbol{k}\)] (a) ,
		(a) -- {(b),(c)},
		(b) -- [scalar, half left, momentum=\(\boldsymbol{p}\)] (c)
	};
	\end{feynman}
	\end{tikzpicture}
	\n
	&= \quad
	\int_0^\Lambda \frac{\mathrm{d}p}{2\pi^2} p^2 P_L (\boldsymbol{p})
	\equiv \sigma^2 (\Lambda).
\end{align}
Thus the lowest order counterterm is $-\sigma^2 (\Lambda)$.
By subtracting tadpole diagrams involving operators that contribute to Eqn.~\eqref{eqn:bias}, we ensure that the expectation value of the biased tracer vanishes at the one-loop level, $\langle \delta_{21} \rangle = 0$. The next counterterm cancels the UV sensitivity of
\begin{align}
    \braket{\left(\delta^2\right)_{\boldsymbol{k}} \delta^{(1)}_{\boldsymbol{q}}} \quad &= \quad 2 \times
	\begin{tikzpicture}[baseline=(current bounding box.center)]
	\begin{feynman}
	\vertex (i);
	\vertex [right=1cm of i, small, blob] (a) {};
	\vertex [above right=1cm of a, dot] (b) {};
	\vertex [below right=1cm of a, dot] (c) {};
	\vertex [right=1cm of b, dot] (d) {};
	\vertex [right=1cm of d] (f);
	\diagram*{
		(i) -- (a) ,
		(a) -- {(b),(c)},
		(b) -- [scalar] (c),
		(b) -- [scalar] (d),
		(d) -- (f),
	};
	\end{feynman}
	\end{tikzpicture}
	\n
	&= \quad
	P_L (\boldsymbol{k}) \int \dbar^3 p \, F_2(\boldsymbol{k}, \boldsymbol{p}) P_L (\boldsymbol{p}) \n
	&= \frac{68}{21} \sigma^2 (\Lambda) P_L (\boldsymbol{k}) .
\end{align}
Since the corresponding counterterm must be proportional to $P_L (\boldsymbol{k})$ when correlated with a factor of $\delta^{(1)}$, the counterterm must be proportional to the operator $\delta$.
Thus, the first few terms of the renormalized $\left(\delta^2\right)_{\boldsymbol{k}}$ are
\begin{equation}
    [\delta^2] = \delta^2 - \sigma^2 (\Lambda) - \frac{68}{21} \sigma^2 (\Lambda) \delta .
\end{equation}
Note that this equation is written in configuration space.
In order to include corrections from higher-order counterterms, we would continue to calculate higher point correlation functions. In this way, we renormalize all the composite operators that appear in our bias expansion. 

The Galilean operator, $\mathcal{G}_2$, is not renormalized at leading order in derivatives~\cite{Assassi:2014fva}.
Even if we had decided to calculate these higher derivative counterterms, the limit $q_i \rightarrow 0$ in the renormalization conditions ensures that these counterterms vanish anyway; thus, within this framework, $\mathcal{G}_2$ is not renormalized.

\section{The 21 cm radiation field in redshift space}
\label{sec:21cm}

The 21\,cm differential brightness temperature is a biased tracer of the underlying matter density and can be written as~\cite{Zaroubi_notes,Furlanetto:2019jso}
\begin{align}
	\delta T_b \approx& \,28 (1+\delta) x_\mathrm{HI} \left( 1 - \frac{T_\mathrm{CMB} (\nu)}{T_\mathrm{spin}} \right) \left(\frac{\Omega_b h^2}{0.0223}\right) \n
	&\times \sqrt{\left(\frac{1+z}{10}\right) \left(\frac{0.24}{\Omega_m}\right)} \left(\frac{H(z)/(1+z)}{\mathrm{d}v_\parallel / \mathrm{d}r_\parallel}\right) \,\mathrm{mK} .
	\label{eqn:dTb}
\end{align}
In this expression, $x_\mathrm{HI}$ is the fraction of hydrogen that is neutral, $T_\mathrm{CMB} (\nu)$ is the CMB brightness temperature, $\Omega_b$ is the baryon density in units of the critical density, $h$ is the Hubble constant in units of $100 \,\mathrm{km}\, \mathrm{s}^{-1} \mathrm{Mpc}^{-1}$, $\Omega_m$ is the mass density in units of the critical density, $H(z)$ is the Hubble expansion at $z$, and $dv_\parallel / dr_\parallel$ is the gradient of the proper velocity along the line of sight. 
The spin temperature $T_\mathrm{spin}$ is defined in terms of the ratio of the occupancy of the spin-1 and spin-0 ground states of hydrogen.
\begin{equation}
    \frac{n_1}{n_0} = 3 \exp(-T_*/T_{\mathrm{spin}}), \label{eqn:spintemp}
\end{equation}
Here, $T_* = 0.0681$ K is the temperature corresponding to the 21\,cm wavelength.
The spin temperature varies throughout space and even throughout individual clumps of neutral hydrogen.
However, since we are studying redshifts well into the EoR, we assume $T_\mathrm{spin} \gg T_\mathrm{CMB}$, so the factor of $\left( 1 - \frac{T_\mathrm{CMB} (\nu)}{T_\mathrm{spin}} \right)$ in Eqn. \eqref{eqn:dTb} becomes saturated and the effect of spatial fluctuations in the spin temperature is negligible.
Henceforth, we neglect spin temperature fluctuations. 
This is a common simplification; however there are also a number of studies that do not assume $T_\mathrm{spin} \gg T_\mathrm{CMB}$ ~\cite{2017MNRAS.464.1365M,2018MNRAS.478.5591M}.
Extending our formalism to higher redshifts relevant for cosmic dawn will require that spin temperature fluctuations be taken into account, and will be the subject of future work.

To be explicit, we define $\delta_{21} = (\delta T_b - \overline{\delta T_b}) / \overline{\delta T_b}$ to be the \textit{fluctuations} in the brightness temperature, and not $\delta T_b$ itself.
Then, by comparing Eqns. \eqref{eqn:bias} and \eqref{eqn:dTb}, we see that measuring the bias coefficients gives us information about the distribution and ionization of the intervening hydrogen, as well as cosmological parameters.

\subsection{From real space to redshift space}
\label{sec:RSDs}

Observations of the 21\,cm radiation field are complicated by the fact that neutral hydrogen has a peculiar velocity which give rise to RSDs. In other words, the measured redshift of the 21\,cm line cannot be attributed purely to the expansion of the Universe.
The distances mapped out from the redshift and ignoring the peculiar velocity form a distorted ``redshift space'' and the coordinates $\boldsymbol{x}_r$ in this space are related to real space coordinates $\boldsymbol{x}$ by
\begin{equation}
	\boldsymbol{x}_r = \boldsymbol{x} + \frac{\boldsymbol{\hat{n}} \cdot \boldsymbol{v}_\mathrm{pec}}{\mathcal{H}} \boldsymbol{\hat{n}} .
\end{equation}
Here, $\boldsymbol{\hat{n}}$ is the line-of-sight direction and $\boldsymbol{v}_\mathrm{pec}$ is the peculiar velocity at the location indicated by the real space coordinate.

The effect of RSDs has been accounted for in effective field theory descriptions of LSS and biased tracers of LSS~\cite{Matsubara:2007wj,Senatore:2014vja,Lewandowski:2015ziq,Perko:2016puo,2018JCAP...12..035D,Donath:2020abv,2020PhRvD.102b3515G,2022MNRAS.513..117L}. RSDs have also been treated perturbatively for the 21\,cm signal~\cite{2012MNRAS.422..926M}, but without including a fully systematic treatment of small-scale nonlinearities as described in the previous section.
To derive the effect of RSDs, we can use the above relationship to transform between the real and redshift space density contrast. 
If the density in real space is $\rho$ and in redshift space is $\rho_r$, then conservation of mass implies $\rho_r (\boldsymbol{x}_r) d^3 x_r = \rho (\boldsymbol{x}) d^3 x$. 
Then, expanding in terms of the density constrast $\delta = \rho/\bar{\rho} - 1$, we find
\begin{equation}
	\delta_r (\boldsymbol{x}_r) = (1 + \delta (\boldsymbol{x})) \left| \frac{\partial \boldsymbol{x}_r}{\partial \boldsymbol{x}} \right|^{-1} -1 .
\end{equation}
Fourier transforming this relation yields
\begin{equation}
    (\delta_r)_{\boldsymbol{k}} = \delta_{\boldsymbol{k}} + \int d^3 x \, e^{-i\boldsymbol{k} \cdot \boldsymbol{x}} \left( \exp\left[ -i \frac{k_\parallel v_\parallel}{\mathcal{H}} \right] - 1 \right) (1 + \delta (\boldsymbol{x})) .
\end{equation}
Here, we have defined $v_\parallel \equiv \boldsymbol{\hat{n}} \cdot \boldsymbol{v}_\mathrm{pec}$ and $k_\parallel \equiv \boldsymbol{\hat{n}} \cdot \boldsymbol{k}$. 

The quantity $k_\parallel v_\parallel$ can be thought of as the rate at which modes of length scale $1/k$ are changing along the line of sight, due to peculiar velocities. 
For the modes of interest, this rate is quite small compared to the expansion rate of the universe because the peculiar velocities are very nonrelativistic, hence $k_\parallel v_\parallel / \mathcal{H}$ is a small quantity. Taylor expanding in this parameter then gives another series expansion,
\begin{align}
	(\delta_r)_{\boldsymbol{k}} =& \delta_{\boldsymbol{k}} 
	-i \frac{k_\parallel}{\mathcal{H}} (v_{\parallel})_{\boldsymbol{k}} 
	-i \frac{k_\parallel}{\mathcal{H}} \left({\delta v_{\parallel}}\right)_{\boldsymbol{k}}
	- \frac{1}{2} \left(\frac{k_\parallel}{\mathcal{H}} \right)^2 \left({v_\parallel^2}\right)_{\boldsymbol{k}} \n
	&- \frac{1}{2} \left(\frac{k_\parallel}{\mathcal{H}} \right)^2 \left({\delta v_\parallel^2}\right)_{\boldsymbol{k}}
	+ \frac{i}{6} \left(\frac{k_\parallel}{\mathcal{H}} \right)^3 \left({v_\parallel^3}\right)_{\boldsymbol{k}}
	+ \cdots
	\label{eqn:RSD_expansion}
\end{align}
We see that we now have new operators that include factors of the velocity field which will also need to be renormalized following the prescription of Sec.~\ref{sec:renorm}.

Note that the $v_\parallel$ that appears in Eqn.~\eqref{eqn:RSD_expansion} is a projection of the \textit{baryon} velocity, while the velocity that appears throughout Section~\ref{sec:review} is the \textit{matter} velocity.
Relative velocities between baryons and dark matter can affect the formation and distribution of the first bound objects, leaving imprints on the matter power spectrum and galaxy bispectrum~\cite{2010PhRvD..82h3520T,2011JCAP...07..018Y}, as well as affecting the Lyman-$\alpha$ forest~\cite{2020PhRvD.102b3515G}, reionization, and the 21\,cm signal~\cite{Munoz:2019fkt,Munoz:2019rhi,Cain:2020npm,Park:2020ydt,2022MNRAS.513..117L}.
However, during the EoR, when the first collapsed objects have already formed, the effect of these relative velocities is negligible~\cite{2011ApJ...730L...1S}.
In the simulations we use, the difference between the two velocities is less than 2\% in the vast majority of the volume and this approximation is also justified and used in other studies~\cite{2012MNRAS.422..926M}. 
Hence, for our purposes, we take the velocity of the neutral hydrogen and matter to be the same and leave the inclusion of a relative velocity term for future study.

\subsection{The effective 21 cm field}
\label{sec:field}

The steps to building up our effective field theory are:
\begin{enumerate}
    \item Use standard perturbation theory to treat the evolution of the matter density field, $\delta$.
    See Section~\ref{sec:spt} for a review of SPT.
    \item Include a bias expansion to write the 21\,cm field $\delta_{21}$ in terms of $\delta$.
    See also Ref.~\cite{Desjacques:2016bnm} for a comprehensive review of cosmological perturbation theory for biased tracers.
    \item Include an RSD expansion in $k_\parallel v_\parallel / \mathcal{H}$, in order to write the redshift space field $\delta_{21,r}$ in terms of the real space field $\delta_{21}$.
    \item Smooth over non-perturbative modes in the field using some wavenumber $\Lambda$ and renormalize the composite operators that appear.
    See Section~\ref{sec:renorm} for more details.
\end{enumerate}
\begin{widetext}
Putting together the bias and RSD expansions, we obtain
\begin{align}
    (\delta_{21,r})_{\boldsymbol{k}} = &b_{1} \delta_{\boldsymbol{k}} - b_{\nabla^2} k^2 \delta_{\boldsymbol{k}} + b_{2} {\left(\delta^2\right)_{\boldsymbol{k}}} + b_{G2} (\mathcal{G}_2)_{\boldsymbol{k}} \n
    &-i \frac{k_\parallel}{\mathcal{H}} \left[ (v_\parallel)_{\boldsymbol{k}} + b_1 ({\delta \,v_\parallel})_{\boldsymbol{k}} - b_{\nabla^2} k^2 ({\delta\, v_\parallel})_{\boldsymbol{k}} \right] - \frac{1}{2} \left(\frac{k_\parallel}{\mathcal{H}} \right)^2 ({v_\parallel^2})_{\boldsymbol{k}} + \cdots
    \label{eqn:d21_bare}
\end{align}
There are terms in Eqn.~\eqref{eqn:d21_bare} that are not multiplied by any bias coefficients; thus, when fitting the theory to data or simulations, the size of these terms cannot be adjusted. 
We have checked that in the linear limit, the bias-independent term in Eqn. \eqref{eqn:d21_bare} resulting from RSDs matches the result used in Ref~\cite{Barkana:2004zy}.
RSDs therefore give rise to a bias-independent contribution to the power spectrum, which can enhance the 21\,cm power spectrum relative to the matter spectrum by a factor of up to $\sim 2$~\cite{Barkana:2004zy}.
The measurability of these contributions depends on redshift~\cite{2015PhRvL.114j1303F} and the angular dependence of these terms can also be used to distinguish the contributions to the 21\,cm power spectrum due to density fluctuations from ionization fluctuations~\cite{2012MNRAS.422..926M}.

We can now apply the procedure outlined in section \ref{sec:renorm} to renormalize the operators appearing in Eqn.~\eqref{eqn:d21_bare}.
As discussed above, $\mathcal{G}_2$ is not renormalized at leading order in derivatives~\cite{Assassi:2014fva}, and furthermore $\left(\delta\, v_\parallel\right)_{\boldsymbol{k}}$ receives no extra counterterms because the momentum $\pi_i \propto (1 + \delta) v_i$ is automatically renormalized through the continuity equation~\cite{Senatore:2014vja,Perko:2016puo}.
\footnotemark[2]
Then for the remaining operators, we find
\begin{align}
	\left[\delta^2\right] &= \delta^2 - \sigma^2 (\Lambda) \left( 1 + \frac{68}{21} \delta + \frac{8126}{2205} \delta^2 + \frac{254}{2205} \mathcal{G}_2 \right) + \cdots \n
	\left[ v_\parallel^2 \right] &= v_\parallel^2 - \mathcal{H}^2 \varsigma^2 (\Lambda) \left[ \frac{1}{3} + \frac{2}{105} \left( 24 + 23 \frac{k_{\parallel}^2}{k^2} \right) \delta + v^{(2)}_{ct} \right] + \cdots
\end{align}
where $\varsigma^2 (\Lambda) = \int_0^\Lambda \frac{\mathrm{d}p}{2\pi^2} P_L (\boldsymbol{p})$ and $v^{(2)}_{ct}$ is given in Fourier space by 
\begin{align}
    \left( v^{(2)}_{ct} \right)_{\boldsymbol{k}} &= \frac{\delta^D (\boldsymbol{q}_1 + \boldsymbol{q}_2 - \boldsymbol{k})}{10290} \left[ 996 + 2041 \left( \frac{q_{1,\parallel}^2}{q_1^2} + \frac{q_{2,\parallel}^2}{q_2^2} \right) - 2142 \frac{q_{1,\parallel} q_{2,\parallel}}{q_1 q_2} \right. \n
    &\left.\qquad\qquad\quad + \frac{\boldsymbol{q}_{1} \cdot \boldsymbol{q}_{2}}{q_1 q_2} \left( 1071 \frac{q_{1,\parallel}^2}{q_1^2} + 1071 \frac{q_{2,\parallel}^2}{q_2^2} - 948 \frac{\boldsymbol{q}_{1} \cdot \boldsymbol{q}_{2}}{q_1 q_2} + 2844 \frac{q_{1,\parallel} q_{2,\parallel}}{q_1 q_2} \right) \right] \delta_{\boldsymbol{q_1}} \delta_{\boldsymbol{q_2}} .
\end{align}
More details about the derivation of the counterterms for $v_\parallel^2$ can be found in Appendix~\ref{sec:v2ct}.
Re-expressing Eqn.~\eqref{eqn:d21_bare} in terms of these renormalized operators, we obtain
\begin{align}
    (\delta_{21,r})_{\boldsymbol{k}} = &b_{1}^{(R)} \delta_{\boldsymbol{k}} - b_{\nabla^2} k^2 \delta_{\boldsymbol{k}} + b_{2}^{(R)} \left[ \delta^2 \right]_{\boldsymbol{k}} + b_{G2}^{(R)} (\mathcal{G}_2)_{\boldsymbol{k}} \n
    &-i \frac{k_\parallel}{\mathcal{H}} \left[ (v_\parallel)_{\boldsymbol{k}} + b_1 (\delta v_\parallel)_{\boldsymbol{k}} - b_{\nabla^2} k^2 (\delta v_\parallel)_{\boldsymbol{k}} \right] - \frac{1}{2} \left(\frac{k_\parallel}{\mathcal{H}} \right)^2 \left[ v_\parallel^2 \right]_{\boldsymbol{k}} + \cdots
    \label{eqn:d21_renorm}
\end{align}
where the renormalized bias coefficients are given by
\begin{align}
    b_1^{(R)} &= b_{1} + \sigma^2 (\Lambda) \left( \frac{34}{21} b_2 \right) - \frac{2}{420} \left( 24 + 23 \frac{k_{\parallel}^2}{k^2} \right) k_\parallel^2 \varsigma^2 (\Lambda) \n
    b_2^{(R)} &= b_{2} + \frac{8126}{2205} \sigma^2(\Lambda) b_2 - \frac{1}{2} k_\parallel^2 \varsigma^2 (\Lambda) v^{(2)}_{ct} \n
    b_{G2}^{(R)} &= b_{G2} + \frac{254}{2205} \sigma^2 (\Lambda) b_2 .
\end{align}
\end{widetext}
In these expansions, we have only gone to second order in fields.
This is because renormalized operators that start at third order in $\delta^{(1)}$ do not contribute to the one-loop power spectrum.
For example, consider the bare operator $\delta^3$.
This operator's only contribution to the one-loop power spectrum is through the correlation function with the linear density field $\braket{(\delta)^3 \delta^{(1)}}$, which begins at one-loop order and has no tree-level component. To build the renormalized operator $[\delta^3]$, the renormalization condition in Eqn. \eqref{eqn:renorm_condition} requires that the correlation function between $[\delta^3]$ and factors of the linear density field equal the tree-level contribution, which is zero.
Hence, these third order operators have no contribution at one-loop order. This is in contrast to the bare operator $\delta^2$, which contributes to the one-loop power spectrum through $\braket{(\delta)^2 \delta^{(2)}}$; the renormalization procedure does not null out this contribution unlike for the case of $\delta^3$.
\footnotetext{To be more explicit, the momentum can be decomposed into gradient and curl components as $\pi^i = a \bar{\rho} \left( \frac{\partial^i}{\partial^2} \pi_s + \epsilon^{ijk} \frac{\partial_j}{\partial^2} \pi_{v,k} \right)$. 
The scalar potential $\pi_s$ is related to the density by the continuity equation, $\pi_s = - \dot{\delta}$, and so receives the same counterterms as $\delta$. 
The vector potential $\pi_v$ does not need to be renormalized since it first appears at third order, which is all we need for the one-loop power spectrum, and receives no counterterms at this order. 
Thus, $\pi_i$ requires no additional counterterms. Since $\pi_i \propto (1 + \delta) v_i$, and $v_i$ is already renormalized through Eqn.\eqref{eqn:EFT_theta}, then $\delta v_i$ also has no additional counterterms.}

Finally, we use the SPT ansatzes for $\delta$ and $\theta$ to write the field in terms of the linear density perturbations $\delta^{(1)}$.
We substitute
\begin{align}
	\delta &= \delta^{(1)} + \delta^{(2)} + \delta^{(3)}, \\
	\delta^2 &= (\delta^{(1)})^2 + 2 \delta^{(1)} \delta^{(2)} , \\
	\mathcal{G}_2 &= \left( \frac{\nabla_i \nabla_j}{\nabla^2} \delta^{(1)} \right)^2 + 2 \left( \frac{\nabla_i \nabla_j}{\nabla^2} \delta^{(1)} \right) \left( \frac{\nabla^i \nabla^j}{\nabla^2} \delta^{(2)} \right) \n
	&\qquad - (\delta^{(1)})^2 - 2 \delta^{(1)} \delta^{(2)} , \\
	v_\parallel &= - \mathcal{H} \frac{\nabla_\parallel}{\nabla^2} (\theta^{(1)} + \theta^{(2)} + \theta^{(3)}), \\
	\delta v_\parallel &= - \mathcal{H} \left( \delta^{(1)} \frac{\nabla_\parallel}{\nabla^2} \theta^{(1)} + \delta^{(1)} \frac{\nabla_\parallel}{\nabla^2} \theta^{(2)} + \delta^{(2)} \frac{\nabla_\parallel}{\nabla^2} \theta^{(1)} \right) , \\
	v_\parallel^2 &= \mathcal{H}^2 \left( \frac{\nabla_\parallel}{\nabla^2} \theta^{(1)} \frac{\nabla_\parallel}{\nabla^2} \theta^{(1)} + 2 \frac{\nabla_\parallel}{\nabla^2} \theta^{(1)} \frac{\nabla_\parallel}{\nabla^2} \theta^{(2)} \right) .
\end{align}
As we explain in the next section, we also use the second-order approximation $\delta^2 = (\delta^{(1)})^2$, as we find the $\delta^{(1)} \delta^{(2)}$ term is very noisy and affects the quality of the fit to simulations.

\section{Fitting to the \thesan simulations}
\label{sec:thesan}

To validate our EFT calculation of the 21\,cm power spectrum, we fit Eqn.~\eqref{eqn:d21_renorm} to simulations, emphasizing that this procedure is performed at the field level.
For this study, we use the recently developed \thesan simulations~\cite{thesan1,thesan2,thesan3}.

\thesan is a new suite of radiation-magneto-hydrodynamical simulations designed to simultaneously capture the complex physics of cosmic reionization and high-redshift galaxy formation. 
These simulations combine a large comoving box size of $95.5$\,Mpc, high resolution (sufficient to model the formation of atomic cooling halos, the smallest structures significantly contributing to the reionization process), a wide range of self-consistent realistic prescriptions for high-redshift physics (built on top of the successful IllustrisTNG galaxy formation model, described in Refs.~\cite{Weinberger+2017,Pillepich+2018}), and the approach to initial conditions production described in Ref.~\cite{Angulo&Pontzen2016}, which significantly reduces the effect of sample variance, increasing the statistical fidelity of the simulations.
As a reminder, all distances and wavenumbers are reported in comoving units.

The simulations are performed using the code \mbox{\textsc{arepo-rt}} \cite{Arepo, ArepoRT}. 
Radiation-magneto-hydrodynamics equations are solved on a mesh, built from a set of mesh-generating points that approximately follow the gas flow as their Voronoi tessellation. 
This approach ensures a natural increase of resolution in the high-density regions, where it is needed. Gravity is instead computed using an hybrid Tree-PM approach, where long-range forces are computed using a particle mesh algorithm and short-range ones are calculated using a hierarchical oct-tree \cite{Barnes&Hut86}.
The photon production rate for star is computed using the BPASS \cite{BPASS2017, BPASS2018} library. 

Among other observables, the \thesan simulations have been shown to reproduce realistic realizations of the reionization history of the Universe, IGM temperature evolution, optical depth to the CMB, $z \geq 6$ UV luminosity function \cite{thesan1}, photo-ionization rate, mean free path of ionising photons, IGM opacity and temperature-density relation \cite{thesan2}.

The different simulations that make up the \thesan suite are described in Ref.~\cite{thesan1} and shown in Figure \ref{fig:slices}.
Here, we briefly summarize the properties of the simulations we include in this study.
\begin{itemize}
    \item \thesan-1: The highest resolution simulation with $2100^3$ dark matter particles of mass $3.12 \times 10^6\,\Msun$ and $2100^3$ gas particles of mass $5.82 \times 10^5\,\Msun$.
    
    \item \thesan-2: A medium resolution simulation with $1050^3$ dark matter particles of mass $2.49 \times 10^7\,\Msun$ and $1050^3$ gas particles of mass $4.66 \times 10^6\,\Msun$.
    This simulation is the same as \thesan-1, but the spatial resolution has been lowered by a factor of 2 (i.e. the particles in the simulations have been coarse-grained to be more massive by a factor of 8).
    
    \item \thesan-\textsc{wc}-2: Same as \thesan-2, but the birth cloud escape fraction is slightly higher to compensate for lower star formation in the medium resolution runs. 
    The total integrated number of photons emitted in \thesan-1 and \thesan-\textsc{wc}-2 are the same.
    
    \item \thesan-\textsc{high}-2: Same as \thesan-2, but using a halo-mass-dependent escape fraction, with only halos \textit{above} $10^{10}\,\Msun$ contributing to reionization.
    
    \item \thesan-\textsc{low}-2: Same as \thesan-2, but using a halo-mass-dependent escape fraction, with only halos \textit{below} $10^{10}\,\Msun$ contributing to reionization.
    
    \item \thesan-\textsc{sdao}-2: Same as \thesan-2, but using a non-standard dark matter model that includes couplings to relativistic particles. 
    The effect of these new interactions is to give rise to sDAOs and cut off the linear matter power spectrum at small scales~\cite{Cyr-Racine:2013fsa}.
    This difference is quantified through a transfer function, which is defined as the square root of the ratio between the DAO matter power spectrum and the standard cold dark matter power spectrum~\cite{Bohr:2020yoe}.
\end{itemize}

For this study, we use each simulation on a $128^3$ grid.
Higher resolutions are not necessary since our methods are only relevant on the largest scales.
To simulate the redshift space distortions, we create a mock observer and adjust the particle data according their peculiar velocities.
Due to the $128^3$ render grid that we use, we can only resolve RSDs corresponding to peculiar velocities greater than $\Delta v = H(z) \Delta r = 39 \,\mbox{km/s} \times \sqrt{\frac{1+z}{7}} \frac{128}{N_\mathrm{pix}}$, where we have used Hubble's law, $\Delta r$ is the smallest change in distance we can resolve on the grid, and $N_\mathrm{pix}$ is the number of pixels along one dimension of the grid.

\subsection{Fitting the bias parameters}
\label{sec:fitting}

\begin{figure}
	\includegraphics{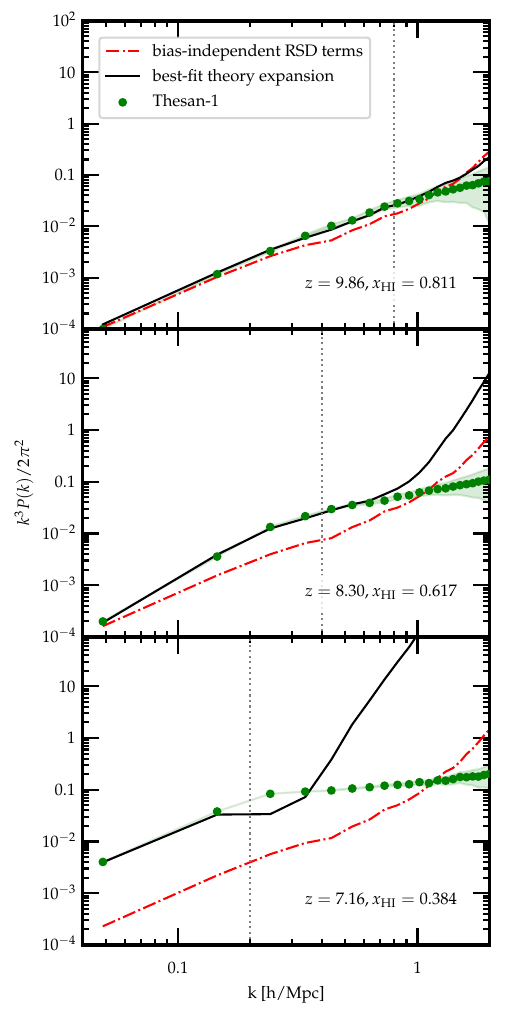}
	\caption{21\,cm power spectra in redshift space at various redshifts and values for the neutral hydrogen fraction.
	Green dots indicate the binned power spectrum from the \thesan-1 simulation; the shaded regions indicate the shot noise error.
	The black dashed line shows the best fit from the effective field theory. 
	The fits are performed at the level of the fields, rather than at the power spectrum level.
	The red dash-dotted line shows the relative contribution of the terms with no bias coefficients. 
	The vertical gray dotted line is at $k_\mathrm{NL}$, which is the maximum wavenumber that we fit up to.
	For earlier redshifts, our theory expansion remains a good fit to \thesan at slightly larger wavenumbers than $k_\mathrm{NL}$, indicating that the 21\,cm intensity field is still perturbative at these times and our theory still has predictive power for smaller scales than the ones that we fit.
	}
	\label{fig:power_spectra}
\end{figure}

To test our perturbative expansion, we use the same method as in Ref.~\cite{McQuinn:2018zwa} and fit to simulations at the level of the fields, instead of the power spectrum itself.
Since the power spectrum has a broad shape and our bias expansion has many parameters, fitting directly at the level of the power spectrum could be subject to overfitting. 
Instead, we fit the \textit{fields} at every mode with wavenumber less than $k_\mathrm{NL}$, where for practical purposes we define $k_\mathrm{NL}$ to be the scale at which the simulated 21~cm field smoothed over $k_\mathrm{NL}$ has a maximum value of $|\delta_\mathrm{sim}| = 0.8$ in redshift space. 
In other words, $k_\mathrm{NL}$ is analogous to the smoothing scale $\Lambda$ we introduced in Section~\ref{sec:eft}.
In principle, one should choose the $\Lambda$ to be much less than $k_\mathrm{NL}$; however, the number of modes available to fit drastically decreases as we lower the wavenumber cutoff, from 6043 modes at $\Lambda = 1.1$ h/Mpc to 7 modes at $\Lambda = 0.1$ h/Mpc.
Hence, we find that using a smaller $\Lambda$ only increases the fit error, without substantially changing the best fit parameters, so we choose to use $\Lambda = k_\mathrm{NL}$.
In addition, one could choose a different criterion with which to define the nonlinear wavenumber, but we find that varying this threshold between 0.6 and 1.0 also does not substantially change the range of wavenumbers that we fit.

In fitting at the field level, we take the error power spectrum,
\begin{equation}
    P_\mathrm{err} (\boldsymbol{k}) = \frac{|(\delta_\mathrm{sim})_{\boldsymbol{k}} - (\delta_\mathrm{EFT})_{\boldsymbol{k}}|^2}{V} ,
\end{equation}
where $V$ is the simulation volume, and minimize the value of $P_\mathrm{err} (\boldsymbol{k})$ over all modes up to $k_\mathrm{NL}$.
We emphasize that this quantity is \textit{not} the same as the error on the power spectrum; it is the power spectrum of errors at the field level.
Thus, our cost function is
\begin{equation}
    \mathcal{A} = \sum_{\boldsymbol{k}} w_{\boldsymbol{k}} P_\mathrm{err} (\boldsymbol{k}),
\end{equation}
where the sum is over every \textit{mode} and not just every wavenumber value, $w_{\boldsymbol{k}}$ is the weight that we assign to each mode and quantifies how we smooth the fields, $\delta_\mathrm{sim}$ is the simulation field we want to fit to, and $\delta_\mathrm{EFT}$ is the theory expansion.
In this study, $\delta_\mathrm{sim}$ describes perturbations to the redshifted 21\,cm brightness temperature, neglecting fluctuations in spin temperature since we are assuming the $T_\mathrm{spin} \gg T_\mathrm{CMB}$ limit, and $\delta_\mathrm{EFT}$ is given by Eqn.~\eqref{eqn:d21_renorm}. 
The error power spectrum is also sometimes referred to as ``stochasticity'' and is commonly used to quantify the error in estimators of fields~\cite{Seljak:2004ni,Bonoli:2008rb,2010PhRvD..82d3515H,2011MNRAS.412..995C,Baldauf:2013hka,Modi:2016dah}.
If we were able to perfectly construct the 21\,cm field using our model, we would expect $P_\mathrm{err} (\boldsymbol{k}) = 0$.
However, there is an irreducible shot noise contribution to the error due to the discreteness of the simulation particles, which is given by~\cite{2018MNRAS.475..676S}
\begin{equation}
    P_\mathrm{shot} = \frac{V}{N_\mathrm{eff}},
    \label{eqn:Pshot}
\end{equation}
where $N_\mathrm{eff}$ is the effective number of neutral hydrogen tracers, given by
\begin{equation}
    N_\mathrm{eff} = \frac{M^2}{\braket{m^2}}.
    \label{eqn:Neff}
\end{equation}
In this expression, $M = \sum_i m_i$ is the total mass of the tracers and $\braket{m^2} = \left( \sum_i m_i^2 \right) / N$ their mean squared mass, with $N$ being the total number of tracers.

Since we are only fitting modes with wavenumber less than $k_\mathrm{NL}$,
\begin{equation}
    w_{\boldsymbol{k}} = \begin{cases} 1, k < k_\mathrm{NL}, \\ 0, k > k_\mathrm{NL}. \end{cases}
\end{equation}
This choice of weights corresponds to performing least-squares regression. Instead of implementing a sharp cutoff such as this, we could choose to fit the simulation using a smoother filter, such as a Gaussian that down-weights the relative importance of modes closer to the nonlinear scale in determining the fit parameters. 
However, we find that the best-fit parameters are robust to the choice of filter, and $\mathcal{A}$ is smaller for the sharp cutoff filter compared to the Gaussian filter by about 25-30\%.

To calculate the operators appearing in Eqn.~\eqref{eqn:d21_renorm}, we take $\delta^{(1)}$ to be the initial conditions of the simulation, which are seeded at $1+z = 50$ when the perturbations on the scales of interest should still be in the linear regime.
We calculate $\delta^{(2)}$ and $\delta^{(3)}$ using equivalent methods from Lagrangian perturbation theory, since the Lagrangian theory displacements are easier to compute~\cite{Scoccimarro:1997gr,Baldauf:2015zga}.
The velocity factors can then be calculated using
\begin{align}
    \theta^{(1)} &= \delta^{(1)} , \nonumber \\
    \theta^{(2)} &= \delta^{(2)} + \frac{2}{7} \mathcal{G}_2^{(2)} , \nonumber \\
    \theta^{(3)} &= \delta^{(3)} + \frac{2}{9} \left[\mathcal{G}_{2,v} + \frac{1}{7} \nabla^2 \left( \nabla_i \phi \frac{\nabla^i}{\nabla^2} \mathcal{G}_2 \right) \right]^{(3)} .
\end{align}
Above, we denote $\mathcal{G}_{2,v} = \nabla_i \nabla_j \phi \nabla^i \nabla^j \phi_v - \nabla^2 \phi \nabla^2 \phi_v$, where $\phi_v = \theta / \nabla^2$ is the velocity potential.
The superscript 3 at the end of the brackets indicates that we are only keeping terms up to third order in $\delta^{(1)}$.
See Appendix~\ref{sec:theta} for a derivation of these relationships.
Finally, we find that using the second-order approximation for the term $\delta^2 = (\delta^{(1)})^2$ leads to a better fit at most redshifts, see Appendix~\ref{sec:d2_or_d3} for details.
Hereafter, we only show fits using the second-order approximation for $\delta^2$.

Figure~\ref{fig:power_spectra} shows the resulting power spectrum that comes from fitting Eqn.~\eqref{eqn:d21_renorm} to the \thesan-1 simulation \emph{fields} at redshifts of $z =$ 9.86, 8.30, and 7.16.
This corresponds to neutral hydrogen fractions of $x_\mathrm{HI} =$ 0.811, 0.617, and 0.384, respectively.
The red dash-dotted lines show the relative contribution of the bias-independent terms that arise from the RSD expansion, namely $-i \frac{k_\parallel}{\mathcal{H}} (v_\parallel)_{\boldsymbol{k}} - \frac{1}{2} \left(\frac{k_\parallel}{\mathcal{H}} \right)^2 ({v_\parallel^2})_{\boldsymbol{k}}$.
In the first panel, which is the earliest redshift at $z=9.86$, we see that although we only fit up to a maximum wavenumber of $k_\mathrm{NL} = 0.8$ h/Mpc, the series expansion power spectrum smoothly diverges from that of the simulation at wavenumbers above $k_\mathrm{NL}$.
We find similar behavior at $z=8.30$, where we instead fit up to $k_\mathrm{NL} = 0.4$ h/Mpc.
In this case, the theory continues to fit the simulation power spectrum up to a wavenumber of about 0.7 h/Mpc.
This is encouraging, as it indicates that our effective theory has some predictive power past the wavenumbers that we fit. 
By the time the simulation reaches $z = 7.16$, reionization has nearly concluded and the the neutral fraction is much smaller compared to the other redshifts we show.
As a result, $\delta_{21,r}$ is becoming nonperturbative even on the largest scales, since the ionized bubbles have grown quite large as well. 
This degrades the quality of the fit in the last panel relative to the previous redshifts.
For this redshift, we use $k_\mathrm{NL} = 0.2$ h/Mpc and the series expansion is not at all predictive above $k_\mathrm{NL}$.

\begin{figure}[h!]
	\includegraphics{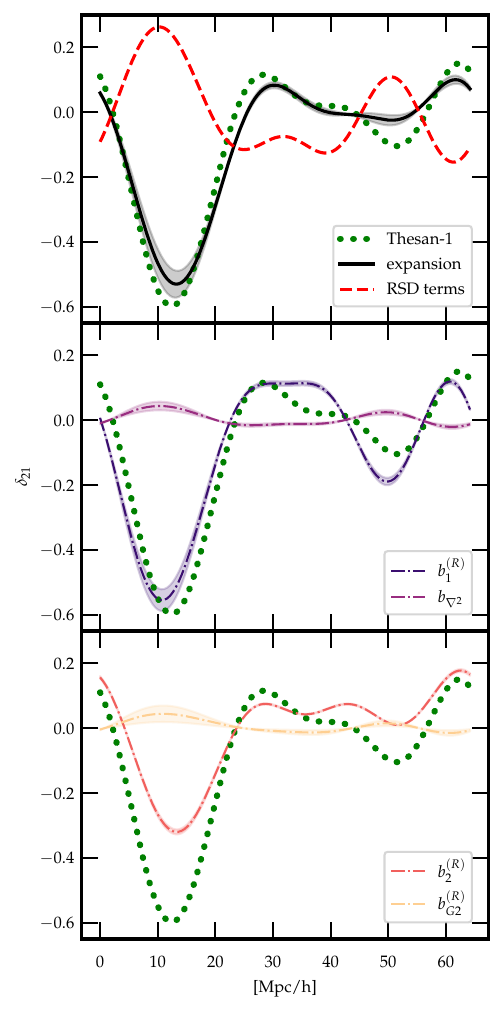}
	\caption{The redshift space 21\,cm differential brightness temperature along a line through the simulation volume at $z = 8.30$, $x_\mathrm{HI} = 0.617$, smoothed over $k_\mathrm{NL} = 0.4$ h/Mpc. 
	The green dots show the signal from the \thesan-1 simulation, the thick black line in the first panel is the best fit theory expansion. 
	For comparison, we also show the contributions of each of the bias parameters to the black line in the other panels, as well as the bias independent contribution (thick red dashed).
	The filled contours show the 68\% confidence intervals on the fitted coefficients.
	}
	\label{fig:1d-slice}
\end{figure}

To see how well the bias expansion fits at the field level, we show in Figure \ref{fig:1d-slice} the fluctuations in the redshift space 21\,cm differential brightness temperature along a line through the simulation volume at $z = 8.30$ and $x_\mathrm{HI} = 0.617$ (in Appendix~\ref{sec:fit_grid}, we show fluctuations in the differential brightness temperature along several lines through the volume).
The fields are smoothed over $k_\mathrm{NL} = 0.4$ h/Mpc.
The green dots show the signal from \thesan-1 and the solid black line in the top panel is the signal from the best fit theory expansion.
We also show the contributions of each term to the best fit theory expansion in the other panels.
Along this particular line, we see that the shapes that dominate the fit are the terms multiplying the $b_1^{(R)}$ coefficient.
Moreover, some of curves show a degree of degeneracy with each other.
Past studies have dealt with such degeneracies using a Gram-Schmidt process to orthogonalize the shapes~\cite{Schmittfull:2018yuk}; we leave an exploration of such a procedure on the 21\,cm field in redshift space to future work. 

To quantify the level of agreement between the simulation and best-fit 21\,cm fields in configuration space, we take the root mean square of fluctuations in the brightness temperature, as well as their difference.
We find the simulation box has a root mean square fluctuation of $0.134$, the best-fit theory field is $0.133$, and their root mean square difference is $0.050$. 
Thus, the disagreement at the \textit{field level} is about $\sim 30\%$. 
We note that the differences appear greatest at the field's extrema.
In the EFT of LSS, comparisons are not typically done at the level of fields in configuration space, but at the level of the power spectrum.
We emphasize that the level of agreement between the power spectra is still percent-level; however, if we were to use our effective field theory description to ``paint on'' the 21\,cm field over a linear initial density field, the root-mean-square difference with the simulated 21\,cm field would likely be degraded at the level of $\sim \mathcal{O}(10\%)$.

\begin{figure}
	\includegraphics{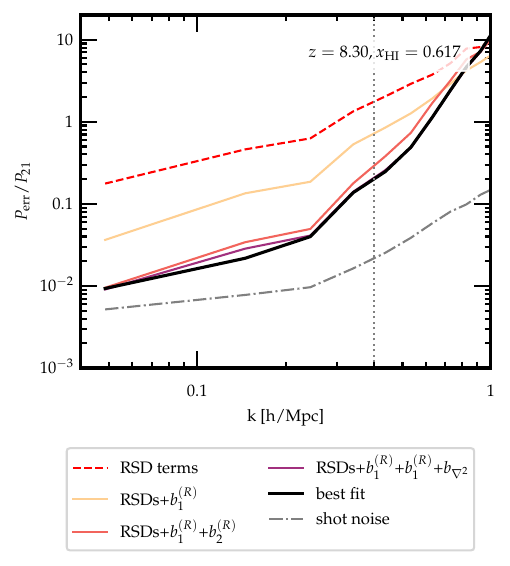}
	\caption{The error power spectrum for the best fit theory field to the \thesan-1 simulation at $z = 8.30$ and $x_\mathrm{HI} = 0.617$.
	Starting with the contribution of the bias-independent RSD terms, the error spectrum is reduced as we add terms one by one, starting with the most dominant terms.
	Shown in the grey dash-dotted line is the contribution from shot noise.
	}
	\label{fig:error}
\end{figure}

We can also assess goodness of fit by looking at the error power spectrum with the best fit coefficients.
Figure~\ref{fig:error} shows the error power spectrum at $z = 8.30$ for the best fit to \thesan-1.
The red dashed line shows the contribution of the bias-independent RSD terms; as the most dominant terms in the theory are added one by one, the error power spectrum is reduced.
For comparison, we also show the curve for shot noise (grey dash-dotted), which is calculated using Eqns.~\eqref{eqn:Pshot} and \eqref{eqn:Neff}.

\section{Discussion}
\label{sec:discussion}

\begin{figure}
	\includegraphics{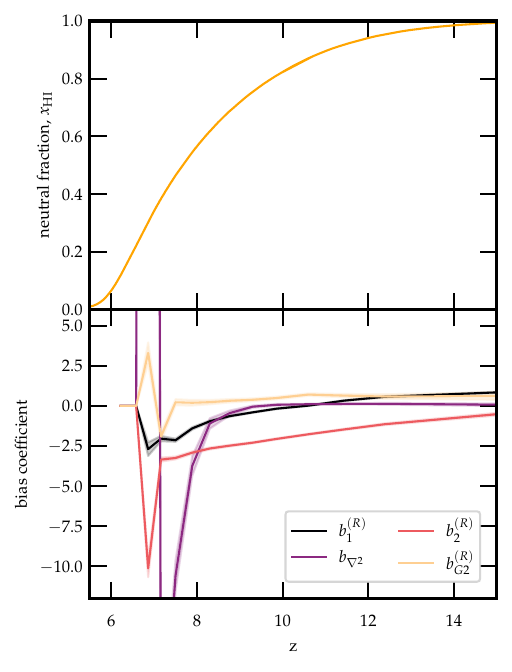}
	\caption{Evolution of the neutral hydrogen fraction and bias coefficients for \thesan-1.
	The top panel shows the reionization history of \thesan-1.
	The bottom panel shows how the bias coefficients change between $z \in [6,15]$.
	The filled contours show the 68\% confidence intervals on the fitted coefficients.
	The evolution of the coefficient is quite smooth down to redshift $z \sim 7.5$, but evolves rapidly thereafter, signalling the breakdown of the perturbation theory.}
	\label{fig:evo}
\end{figure}

Figure~\ref{fig:evo} shows the evolution of the neutral hydrogen fraction and best-fit bias coefficients for \thesan-1.
While the evolution of the bias coefficients is smooth prior to $z \sim 7.5$, at later times the evolution is very rapid, signalling that our perturbative treatment is breaking down as the universe becomes very ionized.

The bias parameters have natural physical interpretations:
\begin{itemize}
    \item $b_1^{(R)}$ is the linear bias and measures how well the 21\,cm field traces the underlying linear matter density.
    
    \item $b_{\nabla^2}$ is related to the effective size of the ionization bubbles $R_\mathrm{eff}$, as argued in Ref.~\cite{McQuinn:2018zwa},
    $$b_{\nabla^2} = \frac{1}{3} b_1^{(R)} R_\mathrm{eff}^2.$$
    As we would expect, this quantity is small at the beginning of reionization, but grows larger with time.
    Once this quantity becomes very large, we expect the ionization field to be quite nonperturbative, hence our formalism will no longer apply.
    
    \item $b_2^{(R)}$ is the quadratic bias, and therefore related to nonlinearities in the 21\,cm field. 
    
    \item $b_{G2}^{(R)}$ is the coefficient for the tidal field, which captures the effects of local anisotropies in the matter field. 
    The length scales of these anisotropies are small compared to the ionization bubbles, hence on the large scales we consider, this term should be subdominant. 
    In addition, the galaxies that source ionizing radiation are highly biased tracers, hence the quadratic bias terms will dominate over the tidal term~\cite{McQuinn:2018zwa}. 
\end{itemize}

\begin{figure*}
	\includegraphics[scale=0.85]{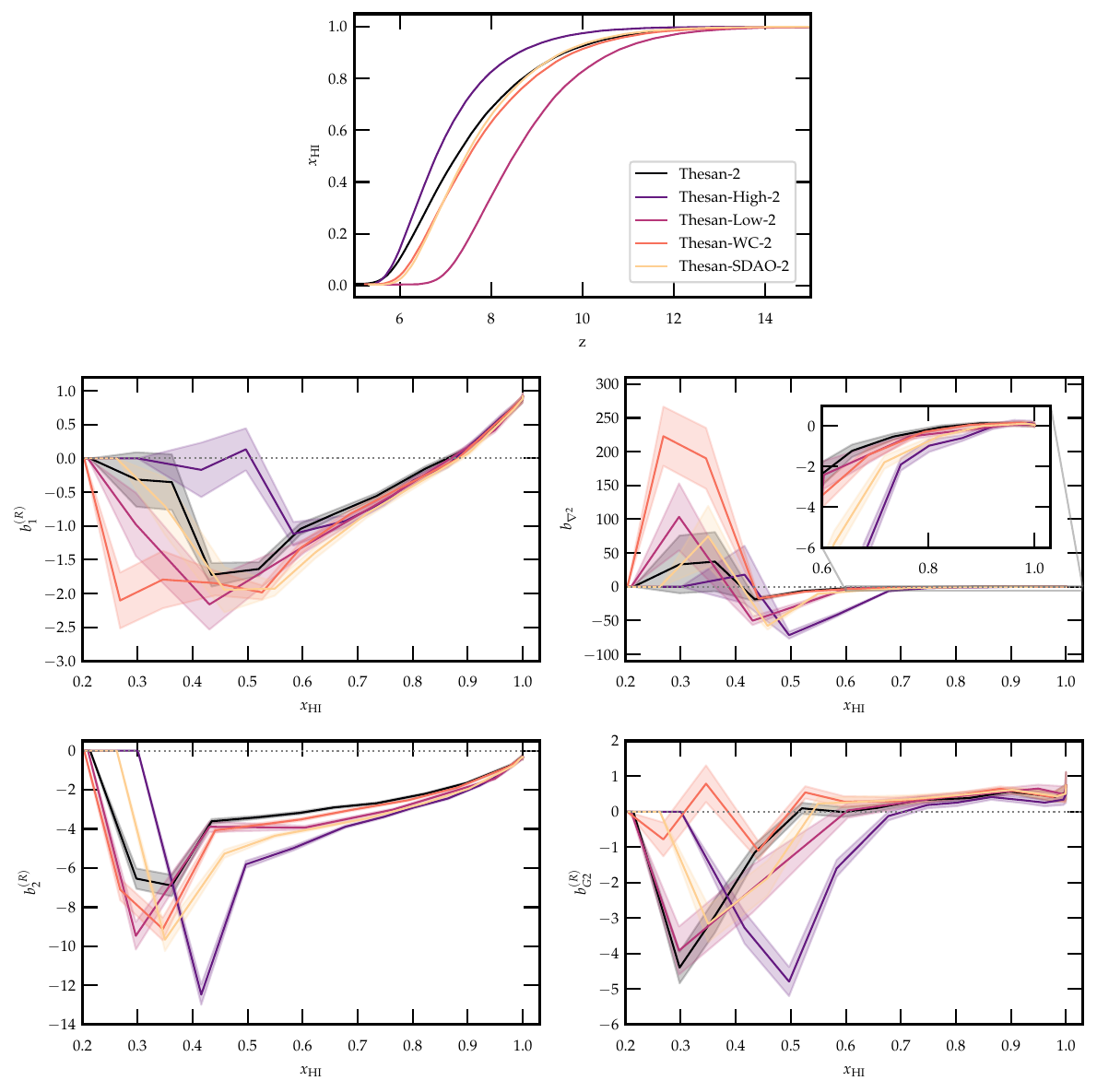}
	\caption{Evolution of the neutral hydrogen fraction and bias coefficients for the various \thesan-2 simulations.
	The filled contours show the 68\% confidence intervals on the fitted coefficients.
	For these fits, we continue to use the approximation $\delta^2 = (\delta^{(1)})^2$, which yields a slightly better fit for many redshifts compared to the third-order expression.
	The top panel shows the neutral fraction as a function of redshift for the different simulations---since the ionization histories vary widely, it is more appropriate to compare simulations at similar values of $x_\mathrm{HI}$, instead of redshift.
	The remaining panels show the best-fit bias coefficients as a function of $x_\mathrm{HI}$.
	At early times, i.e. high neutral fractions, the parameters evolve relatively smoothly; the curves begin to diverge at different values of the neutral fraction, indicating when the perturbative expansion breaks down for each simulation.}
	\label{fig:compare_bias}
\end{figure*}

In order to verify these interpretations of the bias parameters, we can look at how the fitted parameters change between simulations run with different physical models.
Figure~\ref{fig:compare_bias} compares the ionization histories and bias parameters for the five different versions of \thesan-2.
Since the ionization histories can vary widely depending on the physics, fair comparisons of the simulations should be done at similar values of the neutral hydrogen fraction, instead of similar redshifts.
Hence, the remaining panels of Figure~\ref{fig:compare_bias} show the evolution of the bias parameters as a function of $x_\mathrm{HI}$.
At early times and high neutral fractions, the evolution of the bias parameters is relatively smooth.
As reionization progresses and $x_\mathrm{HI}$ decreases, the curves begin to diverge from each other and evolve more rapidly, indicating when the perturbative expansion breaks down for each of the simulations.
At late times, the fitted values of these coefficients should not be trusted; however, it is not surprising that they tend towards zero at the very end of reionization, since the 21\,cm signal should vanish as the neutral fraction goes to zero.

For all of the simulations, $b_{\nabla^2}$ is initially small, then blows up after some critical value for the neutral fraction.
Note that since $b_1^{(R)}$ is negative at the redshifts where significant bubble growth occurs, we expect $b_{\nabla^2}$ to also be negative according to its relation with $R_\mathrm{eff}$.
$b_{\nabla^2}$ diverges earliest for \thesan-\textsc{high}-2 and \thesan-\textsc{sdao}-2, as we would expect since these simulation source the largest bubbles; the coefficients exceed $b_{\nabla^2} < -5$ before reionization has even reached the halfway point.
In contrast, the other three simulations evolve quite similarly at high neutral fractions, with $b_{\nabla^2} > -4$ up to $x_\mathrm{HI} = 0.6$. In addition, the coefficient $b_2^{(R)}$ starts near zero at the beginning of reionization and grows in magnitude with time.
This indicates the growing bias of the signal over time; since $b_2^{(R)}$ is consistently most negative for \thesan-\textsc{high}-2 and thesan-\textsc{sdao}-2 at early times, the 21\,cm signal is more highly biased in these models.
Again, this is not surprising, since reionization is driven by the largest halos in these simulations, and such halos form in the largest overdensities. Finally, $b_{G2}^{(R)}$ is relatively small at early times for all the simulations, with values ranging between about 0 and 0.5 for $x_\mathrm{HI} > 0.7$.
This is in line with our argument that contributions to the 21\,cm power spectrum from anisotropies should be small.

\begin{figure}
	\includegraphics{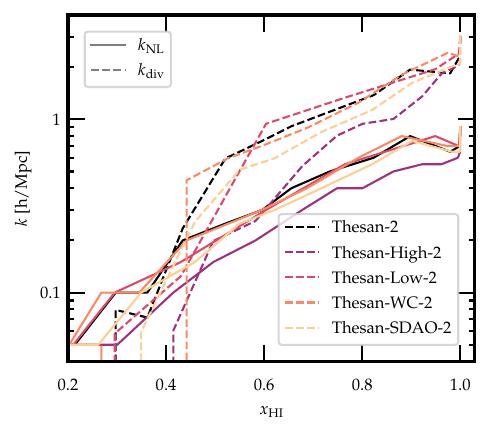}
	\caption{$k_\mathrm{NL}$ (solid lines) and $k_\mathrm{div}$ (dashed lines) as a function of redshift for the various \thesan-2 simulations.
	Regions where $k_\mathrm{div} > k_\mathrm{NL}$ indicate that our theory has predictive power past the wavenumbers that we fit, demonstrating that perturbative methods are valid at high enough values for the neutral fraction, i.e. early enough in the process of reionization.}
	\label{fig:compare_ks}
\end{figure}

To estimate the validity of our EFT methods beyond the range of modes we fit to, we define $k_\mathrm{div}$ as the wavenumber at which the power spectrum of the perturbative expansion with the best-fit bias parameters diverges from the simulation power spectrum by a factor of two.
Figure~\ref{fig:compare_ks} shows $k_\mathrm{NL}$ (dashed lines) and $k_\mathrm{div}$ (solid lines) as a function of $x_\mathrm{HI}$.
At early times or high enough neutral fractions, $k_\mathrm{div} > k_\mathrm{NL}$, which indicates the theory has predictive power even at scales smaller than those that we fit to.
This indicates that our EFT method is valid earlier on in reionization.
Notice that $k_\mathrm{div}$ falls below $k_\mathrm{NL}$ at the highest value of $x_\mathrm{HI}$ for \thesan-\textsc{high}-2 compared to the other simulations, again demonstrating that this simulation field is the least perturbative, due to the large size of the ionized bubbles.

\subsection{Observational limits}
\label{sec:experiment}

\begin{figure}
	\includegraphics{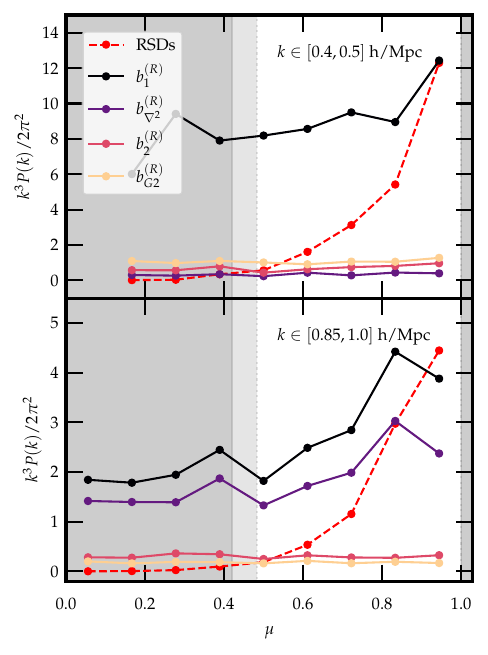}\vspace{-0.4cm}
	\caption{Power in the different operators appearing in the bias expansion, binned over the angle from the line of sight $\mu = \cos\theta$.
	For the top panel, we only average over modes with wavenumbers $k \in [0.4,0.5]$ h/Mpc.
	For the lower panel, we only average over modes with wavenumbers $k \in [0.85,1.0]$ h/Mpc.
	The windows between the light gray (dark gray) regions show the range of $\mu$ that HERA could probe for Band 1 (Band 2), if the foreground wedge could be mitigated.
	We see that all the shapes are relatively flat, except for the term multiplying $b_{\nabla^2}$ on certain scales and the terms appearing due to RSDs.
	The top panel is missing the lowest bin because the simulation did not have modes at those angles for the given range of $k$, due to the lower resolution of the $128^3$ grid used for our analysis.}
	\label{fig:mu_shapes}
\end{figure}

Experiments such as HERA will soon have measurements of the 21\,cm brightness temperature field.
Already, HERA has set upper limits on the 21\,cm power spectra in the spectral windows spanning 117.1--132.6\,MHz (Band 1) and 150.3--167.8\,MHz (Band 2)~\cite{HERA:2021bsv}.
The central frequencies of these bands are 124.8 and 159.0\,MHz, corresponding to redshifts of 10.4 and 7.9, respectively.
HERA and other instruments primarily observe modes along the line of sight; in other words, if we define $\mu = \cos\theta$, where $\theta$ is the angle from the line of sight, then typically instruments probing the 21\,cm power spectrum are most sensitive to larger values of $\mu$. 
Furthermore, due to the chromatic response of interferometers, foregrounds in 21\,cm intensity mapping will leak into the so-called ``foreground wedge"~\cite{2010ApJ...724..526D,2012ApJ...745..176V,2012ApJ...752..137M,Trott:2012md,Thyagarajan:2013eka,Liu:2014bba,Liu:2014yxa}; cuts to avoid foregrounds thus reduce the observable range of $\mu$ even further~\cite{DeBoer:2016tnn}.

If the shapes of the terms that appear in the bias expansion power spectrum look similar across the range of angles that HERA can probe, then varying the bias coefficients of these terms will have the same effect, and a measurement of the power spectrum will not allow us to distinguish the different operators.
To investigate which of the terms in our bias expansion are effectively degenerate, we plot the power in the terms multiplying each bias coefficient in Figure~\ref{fig:mu_shapes} as a function of $\mu$, averaging over $k \in [0.4,0.5]$ h/Mpc and $k \in [0.85,1.0]$ h/Mpc.
The windows between the light gray (dark gray) regions show the range of $\mu$ that HERA could probe in Band 1 (Band 2), if one could mitigate foregrounds within the aforementioned wedge.
From these figures, we see that the power in most of the terms is relatively flat as a function of $\mu$. 
The exceptions are the terms multiplying $b_1$ and $b_{\nabla^2}$ on small scales and the bias-independent terms arising as a result of RSDs.
Thus, without signal contamination from the wedge, HERA could distinguish a few types of shapes in the power spectrum: the relatively flat contribution coming from terms multiplying $b_2^{(R)}$ and $b_{G2}^{(R)}$; the contribution from $b_1^{(R)}$ and $b_{\nabla^2}$; and the contribution from RSDs, which have the strongest scaling with $\mu$ in the relevant range of angles.
In reality, the data cuts that HERA uses to avoid foregrounds limits the observable range to $\mu \gtrsim 0.98$, i.e. nearly directly along the line of sight.
This foreground window may differ between experiments, e.g. LOFAR can probe $\mu \gtrsim 0.97$~\cite{Mertens:2020llj}, and SKA predicts an observable window of $\mu \gtrsim 0.67$~\cite{Wolz:2015sqa}; however, it is clear that the capacity for experiments to distinguish between different shapes in the bias expansions would be greatly improved if one could recover information inside the foreground wedge.

Using the upper limits set by HERA at 95\% confidence level, it is also possible to constrain the possible range of bias parameters. 
A simple method to estimate the constraints would be to set all parameters to zero except the one of interest; we then vary the parameter until we find the values where the power spectrum of the bias expansion lies just under the HERA power spectrum upper limits.
While this is neither the most conservative nor accurate method for constraining the bias parameters, we expect the true values of the bias coefficients to have absolute values much smaller than these limits anyways, since the current upper limits on the power spectrum as measured by HERA are largely set by instrumental systematics and thermal noise~\cite{HERA:2021bsv}.
We find that the estimated HERA constraints on the bias parameters are about one to two orders of magnitude larger than the values fit from simulations. As 21~cm experiments continue to peel away instrumental systematics and take more data, these constraints will shrink and give a more meaningful estimate for the bias parameters.

\section{Conclusions}
\label{sec:conclusion}

In this study, we have incorporated renormalized bias and redshift space distortions into an EFT-inspired description for the 21\,cm brightness temperature.
Using the \thesan simulations, we have shown that these perturbative techniques are valid for describing the behavior of large scales ($k \lesssim 0.8$ h/Mpc) early in reionization ($x_\mathrm{HI} \gtrsim 0.4$). In particular, we can achieve percent-level agreement at the level of the power spectrum and $\mathcal{O}(10\%)$ level agreement at the field level on large scales. We have given physical interpretations for the bias parameters and used simulations run with different physics to test these interpretations.
Since the simulations have very different ionization histories, we have compared them at the same values for ionization and found that the \thesan-\textsc{high}-2 simulation is perturbative for a smaller range of $x_\mathrm{HI}$ due to the larger sizes of the ionization bubbles, while the behavior of \thesan-\textsc{sdao}-2 lies between \thesan-\textsc{high}-2 and the other simulations.

Finally, we have drawn connections between our work and interferometry experiments by showing which shapes in the power spectrum HERA will be able to distinguish in the range of angles that they are sensitive to. We have also estimated how the HERA upper limits on the 21\,cm power spectrum constrain our parameters.

There are many directions that can be taken to further develop this perturbative treatment of the 21\,cm intensity field.
For example, we found some of the terms in the theory are partially degenerate in describing the power spectrum. To reduce such degeneracies and better understand the true degrees of freedom involved, one could apply a method to orthogonalize the shapes appearing in the effective field theory, as was done in Ref.~\cite{Schmittfull:2018yuk}.
In addition, we ignored spin temperature fluctuations in this work; however, this will be an important effect to include if we want to extend this description to describe redshifts where $T_\mathrm{spin} < T_\mathrm{CMB}$.
Such improvements will be critical for using these methods to extract astrophysical and cosmological information from the 21\,cm power spectrum, as we gain important new insights into the EoR in the next few years.

\section*{Acknowledgments}
\label{sec:acknowledgments}

We thank Matt McQuinn for helpful discussions and for sharing his LPT-based code, which we used to calculate the second and third order matter densities.
We also thank Adrian Liu, Juli\'an Mu\~noz, and Ad\'{e}lie Gorce for useful discussions, particularly regarding the connection of this work with HERA.
WQ was supported by the MIT Department of Physics and the National Science Foundation Graduate Research Fellowship under Grant Nos. 1745302 and 2141064.
WQ \& TRS were supported by the U.S. Department of Energy, Office of Science, Office of High Energy Physics of U.S. Department of Energy under grant Contract Number DE-SC0012567. 
KS was supported by a Natural Sciences and Engineering Research Council of Canada (NSERC) Subatomic Physics Discovery Grant and by a Pappalardo Fellowship in the MIT Department of Physics. 
KS and AS acknowledge support for Program numbers HST-HF2-51470.001-A and HST-HF2-51421.001-A provided by NASA through a grant from the Space Telescope Science Institute, which is operated by the Association of Universities for Research in Astronomy, Incorporated, under NASA contract NAS5-26555. 
This research made use of 
\texttt{h5py}~\cite{collette_python_hdf5_2014}, 
\texttt{Jupyter}~\cite{Kluyver2016JupyterN}, 
\texttt{matplotlib}~\cite{Hunter:2007ouj}, 
\texttt{iMinuit}~\cite{James:1975dr,iminuit}, 
\texttt{NumPy}~\cite{Harris:2020xlr}, 
\texttt{SciPy}~\cite{2020NatMe..17..261V}, and 
\texttt{tqdm}~\cite{daCosta-Luis2019}.

\begin{widetext}
\appendix

\section{Counterterms for the $v_\parallel^2$ operator}
\label{sec:v2ct}

The renormalization of $v_\parallel^2$ is very similar to the renormalization of $\delta^2$, as described in Section~\ref{sec:renorm}.
However, the vertices corresponding to the component $\theta^{(n)}_{\boldsymbol{p}}$ fields will come with $G_n$ kernels instead of $F_n$, as well as a factor of $p_\parallel / p = \cos\theta$ for projecting the velocity onto the line of sight.

The zeroth order counterterm is given by
\begin{align}
    \braket{(v_\parallel^2)_{\boldsymbol{k}}} \quad &= \quad
	\begin{tikzpicture}[baseline=(current bounding box.center)]
	\begin{feynman}
	\vertex (i);
	\vertex [right=1cm of i, small, blob] (a) {};
	\vertex [above right=1cm of a, empty dot] (b) {};
	\vertex [below right=1cm of a, empty dot] (c) {};
	\diagram*{
		(i) -- [momentum=\(\boldsymbol{k}\)] (a) ,
		(a) -- {(b),(c)},
		(b) -- [scalar, half left, momentum=\(\boldsymbol{p}\)] (c)
	};
	\end{feynman}
	\end{tikzpicture}
	\n
	&= \quad
	\mathcal{H}^2 \int \dbar^3 p \, P(\boldsymbol{p}) \left( \frac{p_\parallel}{p^2} \right)^2 = \frac{\mathcal{H}^2}{(2\pi)^3} \int d p \, P(\boldsymbol{p}) \times 2\pi \int d \cos\theta \, \cos^2 \theta = \frac{\mathcal{H}^2}{6\pi^2} \int d p \, P(\boldsymbol{p}) .
\end{align}

At $n=1$, there are two identical diagrams that contribute, one each from $\delta^{(1)}$ contracting with one of the component $\theta^{(n)}$ fields.
Each of these diagrams also comes with a symmetry factor of 2, from permuting the two linear legs emerging from $\theta^{(2)}$.
\begin{align}
    \braket{(v_\parallel^2)_{\boldsymbol{q}} \delta^{(1)}_{\boldsymbol{q}}} 
    \quad &= \quad 2 \times
	\begin{tikzpicture}[baseline=(i.base)]
	\begin{feynman}
	\vertex (i);
	\vertex [right=1cm of i, small, blob] (a) {};
	\vertex [above right=1cm of a, empty dot] (b) {};
	\vertex [below right=1cm of a, empty dot] (c) {};
	\vertex [right=1cm of b, dot] (d) {};
	\vertex [right=1cm of d] (f);
	\diagram*{
		(i) -- (a) ,
		(a) -- {(b),(c)},
		(b) -- [scalar, reversed momentum=\(\boldsymbol{p}\)] (c),
		(b) -- [scalar, momentum=\(\boldsymbol{q}\)] (d),
		(d) -- (f),
	};
	\end{feynman}
	\end{tikzpicture}
	\n
    &= - 4 \mathcal{H}^2 P(\boldsymbol{q}) \int \dbar^3 p \, P(\boldsymbol{p}) \frac{p_\parallel}{p^2} \frac{(\boldsymbol{q} - \boldsymbol{p})_\parallel}{(\boldsymbol{q} - \boldsymbol{p})^2} G_2(-\boldsymbol{p}, \boldsymbol{q}) \n
	&= \frac{\mathcal{H}^2}{105} \left( 71 + 23 \frac{q_{\parallel}^2 - q_{\perp}^2}{q^2} \right) P(\boldsymbol{q}) \varsigma^2 (\Lambda)
\end{align}

At $n=2$, there are three diagrams to consider.
\begin{equation}
    \braket{(v_\parallel^2)_{\boldsymbol{q}_1 + \boldsymbol{q}_2} \delta^{(1)}_{\boldsymbol{q}_1}  \delta^{(1)}_{\boldsymbol{q}_2}}
    =
	\begin{tikzpicture}[baseline=(current bounding box.center)]
	\begin{feynman}
	\vertex (i);
	\vertex [right=1cm of i, small, blob] (a) {};
	\vertex [above right=1cm of a, empty dot] (b) {};
	\vertex [below right=1cm of a, empty dot] (c) {};
	\vertex [right=1cm of b, dot] (d) {};
	\vertex [right=1cm of c, dot] (e) {};
	\vertex [right=1cm of d] (f);
	\vertex [right=1cm of e] (g);
	\diagram*{
		(i) -- (a) ,
		(a) -- {(b),(c)},
		(b) -- [scalar] (c),
		(b) -- [scalar] (d),
		(c) -- [scalar] (e),
		(d) -- (f),
		(e) -- (g)
	};
	\end{feynman}
	\end{tikzpicture}
	+ 2 \times
	\begin{tikzpicture}[baseline=(current bounding box.center)]
	\begin{feynman}
	\vertex (i);
	\vertex [right=1cm of i, small, blob] (a) {};
	\vertex [above right=1cm of a, empty dot] (b) {};
	\vertex [below right=1cm of a, empty dot] (c) {};
	\vertex [right=1cm of b, dot] (d) {};
	\vertex [right=1cm of c, dot] (e) {};
	\vertex [right=1cm of d] (f);
	\vertex [right=1cm of e] (g);
	\diagram*{
		(i) -- (a) ,
		(a) -- {(b),(c)},
		(b) -- [scalar] (c),
		(b) -- [scalar] (d),
		(b) -- [scalar] (e),
		(d) -- (f),
		(e) -- (g)
	};
	\end{feynman}
	\end{tikzpicture}
	+
	\begin{tikzpicture}[baseline=(current bounding box.center)]
	\begin{feynman}
	\vertex (i);
	\vertex [right=1cm of i, crossed dot] (a) {};
	\vertex [right=1cm of a, empty dot] (b) {};
	\vertex [above right=1cm of b, dot] (c) {};
	\vertex [below right=1cm of b, dot] (d) {};
	\vertex [right=1cm of c] (f);
	\vertex [right=1cm of d] (g);
	\diagram*{
		(i) -- (a) ,
		(a) -- (b),
		(b) -- [scalar] {(c),(d)},
		(c) -- (f),
		(d) -- (g)
	};
	\end{feynman}
	\end{tikzpicture}
\end{equation}
The crossed dot on the third diagram represents the $n=1$ counterterm.
We label these terms $\mathcal{M}_1$, $\mathcal{M}_2$, and $\mathcal{M}_\mathrm{c.t.}$, respectively.
The first diagram should have a symmetry factor of 8; $2$ from each of the $G_2$ vertices, and another factor of 2 from choosing which of the linear legs contracts with which vertex. 
There will be two diagrams of the second type with identical contributions; each will have a symmetry factor of $3!$ on the $G_3$ kernel. 
The third diagram will again have a symmetry factor of 2 from permuting the external legs or the $G_2$ vertex.
\begin{align}
	\mathcal{M}_1 &= - 2^3 \,\mathcal{H}^2 P(\boldsymbol{q}_1) P(\boldsymbol{q}_2) \int \dbar^3 p \, P(\boldsymbol{p}) G_2 (\boldsymbol{q}_1, -\boldsymbol{p}) G_2 (\boldsymbol{q}_2, \boldsymbol{p}) \frac{(\boldsymbol{q}_1 - \boldsymbol{p})_\parallel}{(\boldsymbol{q}_1 - \boldsymbol{p})^2} \frac{(\boldsymbol{q}_2 + \boldsymbol{p})_\parallel}{(\boldsymbol{q}_2 + \boldsymbol{p})^2} \n
	\mathcal{M}_2 &= 2\times 3!\,  \mathcal{H}^2 P(\boldsymbol{q}_1) P(\boldsymbol{q}_2) \int \dbar^3 p \, P(\boldsymbol{p}) G_3 (\boldsymbol{q}_1, \boldsymbol{q}_2, \boldsymbol{p}) \frac{p_\parallel}{p^2} \frac{(\boldsymbol{q}_1 + \boldsymbol{q}_2 + \boldsymbol{p})_\parallel}{(\boldsymbol{q}_1 + \boldsymbol{q}_2 + \boldsymbol{p})^2} \n
	\mathcal{M}_\mathrm{c.t.} &= - \frac{\mathcal{H}^2}{105} \left( 71 + 23 \frac{(\boldsymbol{q}_1 + \boldsymbol{q}_2)_{\parallel}^2 - (\boldsymbol{q}_1 + \boldsymbol{q}_2)_{\perp}^2}{(\boldsymbol{q}_1 + \boldsymbol{q}_2)^2} \right) \varsigma^2 (\Lambda) \times 2 P(\boldsymbol{q}_1) P(\boldsymbol{q}_2) F_2 (\boldsymbol{q}_1, \boldsymbol{q}_2)
\end{align}
Summing these together, integrating, and keeping only the contribution that is nonzero when $\{q_1, q_2\} \rightarrow 0$, we find
\begin{align}
	\braket{(v_\parallel^2)_{\boldsymbol{q}_1 + \boldsymbol{q}_2} \delta^{(1)}_{\boldsymbol{q}_1}  \delta^{(1)}_{\boldsymbol{q}_2}} 
	&= \mathcal{M}_1 + \mathcal{M}_2 + \mathcal{M}_3 \n
	&= \frac{\mathcal{H}^2}{5145} P(\boldsymbol{q}_1) P(\boldsymbol{q}_2) \varsigma^2 (\Lambda) \left[ 996 + 2041 \left( \frac{q_{1,\parallel}^2}{q_1^2} + \frac{q_{2,\parallel}^2}{q_2^2} \right) - 2142 \frac{q_{1,\parallel} q_{2,\parallel}}{q_1 q_2} \right. \n
	&\qquad\qquad\qquad\qquad\qquad\qquad\left. + \frac{\boldsymbol{q}_{1} \cdot \boldsymbol{q}_{2}}{q_1 q_2} \left( 1071 \frac{q_{1,\parallel}^2}{q_1^2} + 1071 \frac{q_{2,\parallel}^2}{q_2^2} - 948 \frac{\boldsymbol{q}_{1} \cdot \boldsymbol{q}_{2}}{q_1 q_2} + 2844 \frac{q_{1,\parallel} q_{2,\parallel}}{q_1 q_2} \right) \right] .
\end{align}

\section{Perturbative velocity divergences from perturbative densities}
\label{sec:theta}

Here, we derive some relationships between the lowest order $\theta^{(n)}$'s and the operators $\delta^{(1)}$, $\delta^{(2)}$, $\delta^{(3)}$, and $\mathcal{G}_2$.
At first order, we see from the form of the perturbative ansatzes that $\theta^{(1)} = \delta^{(1)}$. 
Deriving the second order relationship is also fairly straightforward.
For compactness, we denote $\int_q = \int \dbar^3 q$.
\begin{align}
\theta^{(2)}_{\boldsymbol{q}} &= \int_{q_1} \int_{q_2} (2\pi)^3 \delta^D (\boldsymbol{q}_1 + \boldsymbol{q}_2 - \boldsymbol{q}) G_2 (\boldsymbol{q}_1, \boldsymbol{q}_2) \delta^{(1)}_{\boldsymbol{q}_1} \delta^{(1)}_{\boldsymbol{q}_2} \n
&= \int_{q_1} \int_{q_2} (2\pi)^3 \delta^D (\boldsymbol{q}_1 + \boldsymbol{q}_2 - \boldsymbol{q}) F_2 (\boldsymbol{q}_1, \boldsymbol{q}_2) \delta^{(1)}_{\boldsymbol{q}_1} \delta^{(1)}_{\boldsymbol{q}_2} \n 
&\qquad\qquad + \frac{2}{7} \int_{q_1} \int_{q_2} (2\pi)^3 \delta^D (\boldsymbol{q}_1 + \boldsymbol{q}_2 - \boldsymbol{q}) \left[ \frac{(\boldsymbol{q}_1 \cdot \boldsymbol{q}_2)^2}{q_1^2 q_2^2} - 1 \right] \delta^{(1)}_{\boldsymbol{q}_1} \delta^{(1)}_{\boldsymbol{q}_2} \n
&= \delta^{(2)}_{\boldsymbol{q}} + \frac{2}{7} \int_{q_1} \int_{q_2} (2\pi)^3 \delta^D (\boldsymbol{q}_1 + \boldsymbol{q}_2 - \boldsymbol{q}) \left[ \frac{(\boldsymbol{q}_1 \cdot \boldsymbol{q}_2)^2}{q_1^2 q_2^2} - 1 \right] \delta^{(1)}_{\boldsymbol{q}_1} \delta^{(1)}_{\boldsymbol{q}_2}
\end{align}
In configuration space, the above expression becomes
$$ \theta^{(2)} = \delta^{(2)} + \frac{2}{7} \mathcal{G}_2 ^{(2)}. $$
Since the perturbative ansatz includes a factor of $- \mathcal{H}$, the velocity is given by $v_i = \frac{\partial_i}{\partial^2} \theta = - \mathcal{H} \frac{\partial_i}{\partial^2} (\theta_1 + \theta_2 + \cdots)$.

The process is analogous for the third order term.
Using the recursive relations given in Refs.~\cite{Goroff:1986ep,Jain:1993jh,Bernardeau:2001qr}, we obtain the following expressions for $\theta^{(3)}$:
\begin{align}
\theta^{(3)}_{\boldsymbol{q}} &= \int_{q_1} \int_{q_2} \int_{q_3} (2\pi)^3 \delta^D (\boldsymbol{q}_1 + \boldsymbol{q}_2 + \boldsymbol{q}_3 - \boldsymbol{q}) G_3 (\boldsymbol{q}_1, \boldsymbol{q}_2, \boldsymbol{q}_3) \delta^{(1)}_{\boldsymbol{q}_1} \delta^{(1)}_{\boldsymbol{q}_2} \delta^{(1)}_{\boldsymbol{q}_3} \\
&= \delta^{(3)}_{\boldsymbol{q}} - \frac{2}{9} \int_{q_1} \int_{q_2} \int_{q_3} (2\pi)^3 \delta^D (\boldsymbol{q}_1 + \boldsymbol{q}_2 + \boldsymbol{q}_3 - \boldsymbol{q}) \left[ \alpha(\boldsymbol{q}_1, \boldsymbol{q}_2 + \boldsymbol{q}_3) F_2(\boldsymbol{q}_2, \boldsymbol{q}_3) - \beta(\boldsymbol{q}_1, \boldsymbol{q}_2 + \boldsymbol{q}_3) G_2(\boldsymbol{q}_2, \boldsymbol{q}_3) \right. \nonumber \\
&\qquad\qquad\qquad\qquad\qquad\qquad\qquad\qquad
\left. + \alpha(\boldsymbol{q}_1 + \boldsymbol{q}_2, \boldsymbol{q}_3) G_2(\boldsymbol{q}_1, \boldsymbol{q}_2) - \beta(\boldsymbol{q}_1 + \boldsymbol{q}_2, \boldsymbol{q}_3) G_2(\boldsymbol{q}_1, \boldsymbol{q}_2) \right] \delta^{(1)}_{\boldsymbol{q}_1} \delta^{(1)}_{\boldsymbol{q}_2} \delta^{(1)}_{\boldsymbol{q}_3}.
\end{align}
The $\alpha(\boldsymbol{q}_1, \boldsymbol{q}_2)$ and $\beta(\boldsymbol{q}_1, \boldsymbol{q}_2)$ kernels are given by
\begin{equation}
    \alpha(\boldsymbol{k}_1, \boldsymbol{k}_2) = \frac{\boldsymbol{k}_1 \cdot (\boldsymbol{k}_1 + \boldsymbol{k}_2)}{k_1^2} , \qquad \beta(\boldsymbol{k}_1, \boldsymbol{k}_2) = \frac{(\boldsymbol{k}_1 + \boldsymbol{k}_2)^2 \boldsymbol{k}_1 \cdot \boldsymbol{k}_2}{2 k_1^2 k_2^2} .
\end{equation}
Introducing the combinations $\boldsymbol{m} = \boldsymbol{q}_2 + \boldsymbol{q}_3$ and $\boldsymbol{n} = \boldsymbol{q}_1 + \boldsymbol{q}_2$, we find
\begin{align}
\theta^{(3)}_{\boldsymbol{q}} &= \delta^{(3)}_{\boldsymbol{q}} - \frac{2}{9} \int_{m} \int_{q_1} (2\pi)^3 \delta^D (\boldsymbol{q}_1 + \boldsymbol{m} - \boldsymbol{q}) \alpha(\boldsymbol{q}_1, \boldsymbol{m}) \int_{q_2} \int_{q_3} (2\pi)^3 \delta^D (\boldsymbol{q}_2 + \boldsymbol{q}_3 - \boldsymbol{m}) F_2(\boldsymbol{q}_2, \boldsymbol{q}_3) \delta^{(1)}_{\boldsymbol{q}_1} \delta^{(1)}_{\boldsymbol{q}_2} \delta^{(1)}_{\boldsymbol{q}_3} \nonumber \\
&\qquad\quad
+ \frac{2}{9} \int_{m} \int_{q_1} (2\pi)^3 \delta^D (\boldsymbol{q}_1 + \boldsymbol{m} - \boldsymbol{q}) \beta(\boldsymbol{q}_1, \boldsymbol{m}) \int_{q_2} \int_{q_3} (2\pi)^3 \delta^D (\boldsymbol{q}_2 + \boldsymbol{q}_3 - \boldsymbol{m}) G_2(\boldsymbol{q}_2, \boldsymbol{q}_3) \delta^{(1)}_{\boldsymbol{q}_1} \delta^{(1)}_{\boldsymbol{q}_2} \delta^{(1)}_{\boldsymbol{q}_3} \nonumber \\
&\qquad\quad
- \frac{2}{9} \int_{n} \int_{q_3} (2\pi)^3 \delta^D (\boldsymbol{n} + \boldsymbol{q}_3 - \boldsymbol{q}) \alpha(\boldsymbol{n}, \boldsymbol{q}_3) \int_{q_1} \int_{q_2} (2\pi)^3 \delta^D (\boldsymbol{q}_1 + \boldsymbol{q}_2 - \boldsymbol{n}) G_2(\boldsymbol{q}_1, \boldsymbol{q}_2) \delta^{(1)}_{\boldsymbol{q}_1} \delta^{(1)}_{\boldsymbol{q}_2} \delta^{(1)}_{\boldsymbol{q}_3} \nonumber \\
&\qquad\quad
+ \frac{2}{9} \int_{n} \int_{q_3} (2\pi)^3 \delta^D (\boldsymbol{n} + \boldsymbol{q}_3 - \boldsymbol{q}) \beta(\boldsymbol{n}, \boldsymbol{q}_3) \int_{q_1} \int_{q_2} (2\pi)^3 \delta^D (\boldsymbol{q}_1 + \boldsymbol{q}_2 - \boldsymbol{n}) G_2(\boldsymbol{q}_1, \boldsymbol{q}_2) \delta^{(1)}_{\boldsymbol{q}_1} \delta^{(1)}_{\boldsymbol{q}_2} \delta^{(1)}_{\boldsymbol{q}_3} \nonumber \\
&= \delta^{(3)}_{\boldsymbol{q}} - \frac{2}{9} \int_{m} \int_{q_1} (2\pi)^3 \delta^D (\boldsymbol{q}_1 + \boldsymbol{m} - \boldsymbol{q}) \alpha(\boldsymbol{q}_1, \boldsymbol{m})  \delta^{(1)}_{\boldsymbol{q}_1} \delta^{(2)}_{\boldsymbol{m}} \nonumber \\
&\qquad\quad
+ \frac{2}{9} \int_{m} \int_{q_1} (2\pi)^3 \delta^D (\boldsymbol{q}_1 + \boldsymbol{m} - \boldsymbol{q}) \beta(\boldsymbol{q}_1, \boldsymbol{m}) \delta^{(1)}_{\boldsymbol{q}_1} \theta^{(2)}_{\boldsymbol{m}} \nonumber \\
&\qquad\quad
- \frac{2}{9} \int_{n} \int_{q_3} (2\pi)^3 \delta^D (\boldsymbol{n} + \boldsymbol{q}_3 - \boldsymbol{q}) \alpha(\boldsymbol{n}, \boldsymbol{q}_3) \theta^{(2)}_{\boldsymbol{n}} \delta^{(1)}_{\boldsymbol{q}_3} \nonumber \\
&\qquad\quad
+ \frac{2}{9} \int_{n} \int_{q_3} (2\pi)^3 \delta^D (\boldsymbol{n} + \boldsymbol{q}_3 - \boldsymbol{q}) \beta(\boldsymbol{n}, \boldsymbol{q}_3) \theta^{(2)}_{\boldsymbol{n}} \delta^{(1)}_{\boldsymbol{q}_3}.
\end{align}
Upon relabeling the wavenumber that is being integrated over, this expression becomes:
\begin{align}
\theta^{(3)}_{\boldsymbol{q}} &= \delta^{(3)}_{\boldsymbol{q}} - \frac{2}{9} \int_{q_1} \int_{q_2} (2\pi)^3 \delta^D (\boldsymbol{q}_1 + \boldsymbol{q}_2 - \boldsymbol{q}) \left\{ \alpha(\boldsymbol{q}_1, \boldsymbol{q}_2) \delta^{(1)}_{\boldsymbol{q}_1} \delta^{(2)}_{\boldsymbol{q}_2} + \left[ \alpha(\boldsymbol{q}_2, \boldsymbol{q}_1) - \beta(\boldsymbol{q}_1, \boldsymbol{q}_2) - \beta(\boldsymbol{q}_2, \boldsymbol{q}_1) \right] \delta^{(1)}_{\boldsymbol{q}_1} \theta^{(2)}_{\boldsymbol{q}_2} \right\} \nonumber \\
&= \delta^{(3)}_{\boldsymbol{q}} - \frac{2}{9} \int_{q_1} \int_{q_2} (2\pi)^3 \delta^D (\boldsymbol{q}_1 + \boldsymbol{q}_2 - \boldsymbol{q}) \left[ \alpha(\boldsymbol{q}_1, \boldsymbol{q}_2) - \alpha(\boldsymbol{q}_2, \boldsymbol{q}_1) + \beta(\boldsymbol{q}_1, \boldsymbol{q}_2) + \beta(\boldsymbol{q}_2, \boldsymbol{q}_1) \right] \delta^{(1)}_{\boldsymbol{q}_1} \delta^{(2)}_{\boldsymbol{q}_2} \nonumber \\
&\qquad\quad - \frac{2}{9} \int_{q_1} \int_{q_2} (2\pi)^3 \delta^D (\boldsymbol{q}_1 + \boldsymbol{q}_2 - \boldsymbol{q}) \left[ \alpha(\boldsymbol{q}_2, \boldsymbol{q}_1) - \beta(\boldsymbol{q}_1, \boldsymbol{q}_2) - \beta(\boldsymbol{q}_2, \boldsymbol{q}_1) \right] \left( \theta^{(1)}_{\boldsymbol{q}_1} \delta^{(2)}_{\boldsymbol{q}_2} + \delta^{(1)}_{\boldsymbol{q}_1} \theta^{(2)}_{\boldsymbol{q}_2} \right). \nonumber \\
\end{align}
Since the choice of momentum labeling was arbitrary, the expressions should be symmetrized over $q_1$ and $q_2$.
\begin{align}
\theta^{(3)}_{\boldsymbol{q}} &= \delta^{(3)}_{\boldsymbol{q}} - \frac{2}{9} \int_{q_1} \int_{q_2} (2\pi)^3 \delta^D (\boldsymbol{q}_1 + \boldsymbol{q}_2 - \boldsymbol{q}) \beta(\boldsymbol{q}_1, \boldsymbol{q}_2) \times 2 \delta^{(1)}_{\boldsymbol{q}_1} \delta^{(2)}_{\boldsymbol{q}_2} \nonumber \\
&\qquad\quad - \frac{2}{9} \int_{q_1} \int_{q_2} (2\pi)^3 \delta^D (\boldsymbol{q}_1 + \boldsymbol{q}_2 - \boldsymbol{q}) \left[ \frac{\alpha(\boldsymbol{q}_1, \boldsymbol{q}_2) + \alpha(\boldsymbol{q}_2, \boldsymbol{q}_1)}{2} - 2 \beta(\boldsymbol{q}_1, \boldsymbol{q}_2) \right] \left( \theta^{(1)}_{\boldsymbol{q}_1} \delta^{(2)}_{\boldsymbol{q}_2} + \delta^{(1)}_{\boldsymbol{q}_1} \theta^{(2)}_{\boldsymbol{q}_2} \right) \nonumber \\
&= \delta^{(3)}_{\boldsymbol{q}} - \frac{2}{9} \int_{q_1} \int_{q_2} (2\pi)^3 \delta^D (\boldsymbol{q}_1 + \boldsymbol{q}_2 - \boldsymbol{q}) \beta(\boldsymbol{q}_1, \boldsymbol{q}_2) \times 2 \delta^{(1)}_{\boldsymbol{q}_1} \delta^{(2)}_{\boldsymbol{q}_2} \nonumber \\
&\qquad\quad + \frac{2}{9} \int_{q_1} \int_{q_2} (2\pi)^3 \delta^D (\boldsymbol{q}_1 + \boldsymbol{q}_2 - \boldsymbol{q}) \left[ \frac{(\boldsymbol{q}_1 \cdot \boldsymbol{q}_2)^2}{q_1^2 q_2^2} - 1 + \beta(\boldsymbol{q}_1, \boldsymbol{q}_2) \right] \left( \theta^{(1)}_{\boldsymbol{q}_1} \delta^{(2)}_{\boldsymbol{q}_2} + \delta^{(1)}_{\boldsymbol{q}_1} \theta^{(2)}_{\boldsymbol{q}_2} \right) \nonumber \\
&= \delta^{(3)}_{\boldsymbol{q}} + \frac{2}{9} \left(\mathcal{G}_{2,v}^{(3)}\right)_{\boldsymbol{q}} - \frac{2}{9} \int_{q_1} \int_{q_2} (2\pi)^3 \delta^D (\boldsymbol{q}_1 + \boldsymbol{q}_2 - \boldsymbol{q}) \beta(\boldsymbol{q}_1, \boldsymbol{q}_2) \times 2 \delta^{(1)}_{\boldsymbol{q}_1} \delta^{(2)}_{\boldsymbol{q}_2} \nonumber \\
&\qquad\qquad\qquad + \frac{2}{9} \int_{q_1} \int_{q_2} (2\pi)^3 \delta^D (\boldsymbol{q}_1 + \boldsymbol{q}_2 - \boldsymbol{q}) \beta(\boldsymbol{q}_1, \boldsymbol{q}_2) \left( \theta^{(1)}_{\boldsymbol{q}_1} \delta^{(2)}_{\boldsymbol{q}_2} + \delta^{(1)}_{\boldsymbol{q}_1} \theta^{(2)}_{\boldsymbol{q}_2} \right) \nonumber \\
\end{align}
As a reminder, we denote $\mathcal{G}_{2,v} = \nabla_i \nabla_j \phi \nabla^i \nabla^j \phi_v - \nabla^2 \phi \nabla^2 \phi_v$, where $\phi_v = \theta / \nabla^2$ is the velocity potential.
Then in configuration space, we can write this as
\begin{align}
    \theta_3 &= \delta_3 + \frac{2}{9} \left[\mathcal{G}_{2,v} + \frac{\nabla^2}{2} \left( \nabla_i \phi \nabla^i \phi_v - \nabla_i \phi \nabla^i \phi \right) \right]_3 \nonumber \\
    &= \delta_3 + \frac{2}{9} \left[\mathcal{G}_{2,v} + \frac{1}{7} \nabla^2 \left( \nabla_i \phi \frac{\nabla^i}{\nabla^2} \mathcal{G}_2 \right) \right]_3
\end{align}
The subscript 3 at the end of the brackets indicates that we are only keeping terms up to third order in $\delta_1$.

\section{Noisiness of the $\delta^2$ operator}
\label{sec:d2_or_d3}

\begin{figure}
	\includegraphics{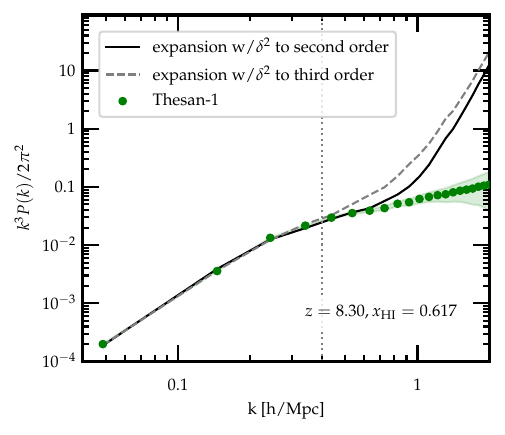}
	\caption{The 21\,cm power spectrum at $z=8.30$ ($x_\mathrm{HI} = 0.617$).
	Green dots indicate the binned power spectrum from the \thesan-1 simulation; the shaded regions indicate the shot noise error.
	The grey dashed line is the theory expansion fit to the simulations using the third-order expression for $\delta^2$, while the black dashed line shows the best fit from the effective field theory, using the approximation $\delta^2 = (\delta^{(1)})^2$.
	The vertical dotted line shows $k_\mathrm{NL}$, the maximum wavenumber that we fit up to.
	Due to the noisiness of the $\delta^2$ operator, we find that the lower-order approximation provides a slightly better fit to the simulation than the full expression, and has greater predictive power for small, mildly nonlinear scales.
	}
	\label{fig:power_spectrum_d2}
\end{figure}

When fitting our theoretical expansion for the 21\,cm field to simulations, we find that using the second-order approximation for the term $\delta^2 = (\delta^{(1)})^2$ leads to a better fit at most redshifts compared to using the full third-order expression.
An example of this is shown in Figure~\ref{fig:power_spectrum_d2}.
The grey dashed line shows the best fit using the third-order expression; the black line uses the second-order expression; the vertical dotted line shows $k_\mathrm{NL}$, the maximum wavenumber that we fit up to.
Not only is the line with the second-order approximation a better fit, it has better predictive power at wavenumbers above those that we fit to.

The degradation of the fit as we go to higher order can be attributed to the fact that $\delta^2$ captures the effects of nonlinear bias and therefore has a large shot noise contribution; including higher order terms then makes this expansion a noisier template when fitting to the simulation field~\cite{McQuinn:2018zwa}.
Hence, the main results of this study use the second-order approximation.

\section{Best fit at the field level}
\label{sec:fit_grid}

\begin{figure*}
	\includegraphics{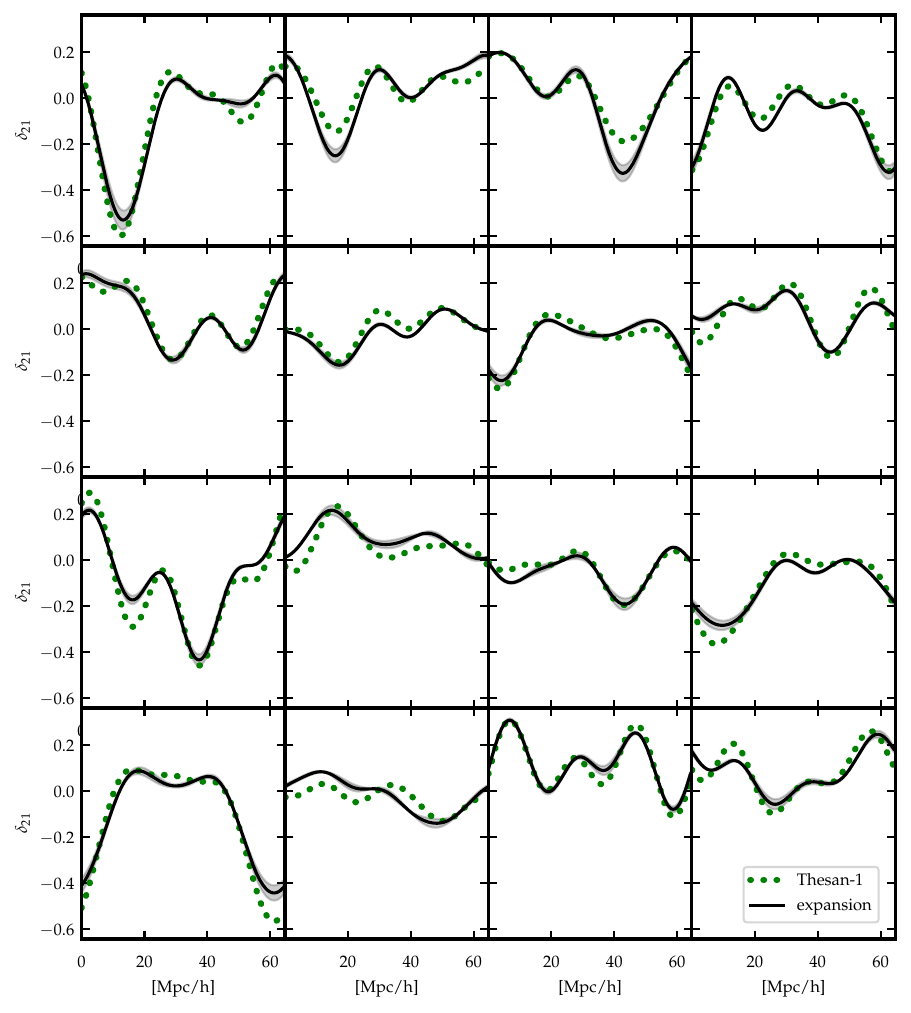}
	\caption{Examples of fluctuations in the redshift space 21\,cm differential brightness temperature along evenly spaced lines through the simulation volume at $z = 8.30$, $x_\mathrm{HI} = 0.617$, smoothed over $k_\mathrm{NL} = 0.4$ h/Mpc. 
	The green dots show the signal from the \thesan-1 simulation and the thick black line in the first panel is the best fit theory expansion.
	The filled contours show the 68\% confidence intervals on the best fit curves.
	}
	\label{fig:1d-slice-grid}
\end{figure*}

Figure~\ref{fig:1d-slice-grid} shows examples of fluctuations in the redshift space 21\,cm differential brightness temperature along several different lines through the simulation volume at $z = 8.30$, $x_\mathrm{HI} = 0.617$, smoothed over $k_\mathrm{NL} = 0.4$ h/Mpc. 
The lines are chosen to be evenly spaced along the $x$ and $y$ coordinates of the simulation volume.

\end{widetext}
\bibliography{21cmEFT}
\end{document}